\documentclass[twocolumn]{aastex62}
\usepackage{dcolumn}
\usepackage{multirow}
\usepackage{amsmath}
\usepackage{hyperref}

\newcolumntype{j}[1]{D{.}{.}{#1}}

\graphicspath{{./}{figures/}{phases/}}
\newcommand{\CISS}{\cass{}/ISS}
\newcommand{\cass}{\emph{Cassini}}
\newcommand{\parange}{14$\degr{}$--24$\degr{}$}
\newcommand{\slices}{6, 4, 6, and 6}

\shorttitle{Icy Terrestrial Phase Curves}
\shortauthors{Mayorga, L.~C. et al.}

\begin{document}

\title{Reflected Light Observations of the Galilean Satellites from Cassini: a testbed for cold terrestrial exoplanets}
\author[0000-0002-4321-4581]{L.~C. Mayorga}
\affiliation{Center for Astrophysics \textbar{} Harvard \& Smithsonian, 60 Garden St, Cambridge, MA 02138, USA}
\affiliation{Johns Hopkins University Applied Physics Laboratory, 11100 Johns Hopkins Rd, Laurel MD 20723, USA}
\author{David Charbonneau}
\affiliation{Center for Astrophysics \textbar{} Harvard \& Smithsonian, 60 Garden St, Cambridge, MA 02138, USA}
\author{D.~P. Thorngren}
\affiliation{Department of Physics, University of California, Santa Cruz, Santa Cruz, CA 95064, USA}
\affiliation{Institut de Recherche sur les Exoplan\`etes, Universit\'e de Montr\'eal, Canada}

\begin{abstract}
For terrestrial exoplanets with thin atmospheres or no atmospheres, the surface contributes light to the reflected light signal of the planet. Measurement of the variety of disk-integrated brightnesses of bodies in the Solar System and the variation with illumination and wavelength is essential for both planning imaging observations of directly imaged exoplanets and interpreting the eventual datasets. Here we measure the change in brightness of the Galilean satellites as a function of planetocentric longitude, illumination phase angle, and wavelength. The data span a range of wavelengths from 400--950~nm and predominantly phase angles from 0$\degr{}$--25$\degr{}$, with some constraining observations near 60$\degr{}$--140$\degr{}$. Despite the similarity in size and density between the moons, surface inhomogeneities result in significant changes in the disk-integrated reflectivity with planetocentric longitude and phase angle. We find that these changes are sufficient to determine the rotational periods of the moon. We also find that at low phase angles the surface can produce reflectivity variations of 8--36\% and the limited high phase angle observations suggest variations will have proportionally larger amplitudes at higher phase angles. Additionally, all the Galilean satellites are darker than predicted by an idealized Lambertian model at the phases most likely to be observed by direct-imaging missions. If Earth-size exoplanets have surfaces similar to that of the Galilean moons, we find that future direct imaging missions will need to achieve precisions of less than 0.1\,ppb. Should the necessary precision be achieved, future exoplanet observations could exploit similar observation schemes to deduce surface variations, determine rotation periods, and potentially infer surface composition.

\end{abstract}

\keywords{stars: planetary systems --- planets and satellites: terrestrial planets and surfaces --- planets and satellites: moons --- techniques: photometric, Exoplanet surface characteristics (496), Exoplanet atmospheres (487), Galilean satellites (627)}

\section{Introduction}
Direct imaging missions are the only proposed technology with the potential to characterize exo-Earths in the habitable zones of nearby solar-type stars in reflected light \citep{Strategy2018}. However, it is unlikely that we will ever be able to angularly resolve the disks of these planets to observe the underlying processes and phenomena that shape the atmosphere and surface. To understand how these sub-resolution processes shape the hemisphere-averaged signal we observe, we look to the solar system. Unresolved observation of objects in the solar system are the only data sources that can provide the necessary high resolution ground truth to connect local variations in reflectivity and color to disk-integrated observations, as has been done for the gas giants and other solar system bodies with thick atmospheres \citep{Mayorga2016, Dyudina2005, Dyudina2016} and the Earth. Modelling efforts of Earth as an exoplanet have been greatly aided by NASA's \emph{EPOXI} mission and have focused on surface features like oceans \citep{Cowan2009,Robinson2010,Robinson2011}, surface color and geography  \citep{Cowan2011,Fujii2010,Fujii2011,Fujii2014,Fujii2017,Livengood2011,Hegde2013}, and potential observational techniques and classification schemes \citep[][and references therein]{Crow2011, Cowan2017a, Fujii2018}. Furthermore, the Deep Space Climate Observatory (\emph{DSCOVR}) has also been used to determine the reflected light signal of Earth as an exoplanet \citep{Jiang2018}.

Reflected light can be a powerful tool in the characterization of planet atmospheres and future large-scale missions, such as \emph{Large UV/Optical/IR Surveyor} \citep[\emph{LUVOIR,}][]{LUVOIRInterim}, and the \emph{Habitable Exoplanet Observatory} \citep[\emph{HabEx},][]{HabEx} will be poised to observe exoplanets in reflected light. However, terrestrial planets pose a much larger observational challenge than giant planets. The surface and atmosphere interaction, evolution, and the additional difficulty in discovering and characterizing them increase their complexity over that of their larger siblings. Until such a time that many more planets are available for study, it is critical that we prepare for what these worlds might hold. 

Through a combination of \replaced{technology}{technological} improvements, terrestrial planets will become more accessible to transit method phase curve studies and direct imaging \added{studies}. With the Solar System as an example, we can expect a range of atmospheric thicknesses from a Venusian atmosphere to a complete lack of atmosphere. For thin atmospheres, like our own, to atmosphereless bodies, the surface is a critical component in understanding the observations. 
The surfaces of the Galilean satellites may serve as analogs to those of cold terrestrial exoplanets such as the TRAPPIST-1~f, g, and h \citep{Gillon2017, Grimm2018} and LHS~1140~b \citep{Dittmann2017, Ment2019}. Despite having similar radii, densities, and equilibrium temperatures, each moon is unique, and collectively they represent a range of potential surfaces that exoplanets could exhibit. Additionally, their interactions with Jupiter's charged particle environment and their responses to tidal stresses have shaped their surfaces, interiors, and any tenuous atmosphere, resulting in complex histories.  All the moons are in slightly eccentric orbits and synchronously rotating.

While there is a great deal of published photometry of the Galilean satellites, both from Earth and from the vicinity of Jupiter, such observations have been little applied to the understanding of exoplanet surfaces or cast in the terms commonly employed in exoplanet observations \citep[see][]{Fujii2014}. From the ground, observations of outer solar system objects are limited to nearly full phase perspectives e.g for Jupiter the phase angles observed are $<$12$\degr{}$. To achieve the full range of phase angles necessary for exoplanet comparison, we require space based observations beyond the orbit of Jupiter. From Earth, rotational light curves have been studied across the spectrum. In the UV, with \emph{Hubble}, the \emph{International Ultraviolet Explorer}, and \emph{Galileo}, \citet{Hendrix2005} measured the disk-integrated rotational and solar phase curves for Europa, Ganymede, and Callisto. Beyond Earth, \emph{Voyager} (and complementary ground-based observations) enabled the measurements of \textless600~nm rotational and solar phase curves for Io \citep{Simonelli1984, Simonelli1986a, Simonelli1986b, Simonelli1986c, Simonelli1988} and the icy satellites \citep{Buratti1983, Buratti1991, Buratti1995}. \emph{Galileo}, which was limited to phase angles of 4\degr{}--14\degr{} and 71\degr{}--86\degr{}, revealed that Io's color and albedo patterns across the surface change dramatically with phase \citep{Simonelli2001}. \citet{Brown2003} and \citet{McCord2004} used the Visual and Infrared Spectrometer on \cass{} to measured the solar phase curve of Europa, Ganymede, and Callisto. 

Here, we expand on prior work by using the Imaging Science Subsystems (ISS) instrument on \cass{}, which observed solar phase angles from 0\degr{}--25\degr{} and approximately 60\degr{}--140\degr{}. Prior observations of the Galilean satellites have enabled a detailed characterization of their surfaces, but here we treat them as point sources, in the same manner that exoplanets will be studied in the future with reflected light. Since the moons are resolved, we can disentangle surface reflectivity variations from illumination effects. For direct imaging missions, full phase observations are impossible and it is important to consider partial phase geometries. Our observations allow us to study the light curves of the Galilean satellites from an exoplanet perspective in an unresolved way to mimic the kinds of observations that will be accessible to us in the future. 

In \autoref{sec:methods}, we introduce the \CISS{} instrument and the data selection, reduction, photometric analyses we undertook to measure the brightnesses of each moon. In \autoref{sec:model}, \added{we} discuss the rotational brightness modulation model and the phase curve fitting analysis conducted. In \autoref{sec:Results}, we present the light curves, brightness maps, and color variations as deduced from unresolved photometric observations as conducted for exoplanets. In \autoref{sec:impact}, we discuss the implications of this work on the future of terrestrial exoplanet study and define requirements for future direct imaging missions before concluding in \autoref{sec:conc}.

\section{Methods}
\label{sec:methods}
While on route to Saturn, \CISS{} took tens of thousands of images of Jupiter during a flyby spanning from 2000 October to 2001 March. The two cameras, the Wide Angle Camera (WAC) and the Narrow Angle Camera (NAC), had identical 1,024 by 1,024 pixel CCD detectors, but the resulting fields-of-view (FOV) for the WAC and NAC were 3\fdg5 and 0\fdg35, respectively. Each camera had two filter wheels for selecting filter combinations of which one filter on each wheel was a clear filter. The NAC had 24 filters and the WAC had 18 filters and they shared 15 filters, albeit with slightly differing spectral transmissions due to the differing optics of the cameras (see \citet{Porco2004}).

\subsection{Data Selection and Reduction}
The FOV of the WAC was sufficiently wide to often capture the Galilean satellites in images that were targeting Jupiter. To identify these images, we examine the index files associated with each \CISS{} archive volume hosted by the Planetary Data System Cartography and Imaging Science Node\footnote{https://pds-imaging.jpl.nasa.gov/volumes/iss.html}. From these index files, we extract the image times, their summation type (binning in 1$\times$1, 2$\times$2, or 4$\times$4), the image number, the distance of Jupiter from the Sun, and the location of the image in the \CISS{} volume directories. We then use SPICE \citep[(Spacecraft Planet Instrument Camera matrix Events),][]{Acton2018}, which has been ported to \verb|Python| as \verb|SpiceyPy| \citep{SpiceyPy}, to track observation geometry and events to determine: (1) spacecraft location and orientation; (2) target location, shape, and size; and (3) events on the spacecraft or the ground that might affect the interpretation of science observations \citetext{SPICE Overview Tutorial 3}. 
While \CISS{} took over 16,000 images of Jupiter, there are only roughly 11,000 images where SPICE predicts that any of the Galilean satellites are visible. Because the moons were not the primary targets in the majority of the images, roughly 60\% of these are removed due to a combination of factors discussed in \autoref{ssec:phot}.

We reduce the images containing the satellites using \verb|CISSCALv3.9|, the \CISS{} reduction pipeline \citep{Knowles2016}. We followed the steps outlined in \citet{Mayorga2016}, and as outlined by \citet{West2010} and \citet{Knowles2016}.

\subsection{Photometry}
\label{ssec:phot}
\CISS{} images are typically compressed, either with a lossless or lossy option. We move forward with only those images that were not compressed with the lossy option, a space and bandwidth saving mode that was eventually discontinued. In the lossless images, which make up roughly 75\% of the data, the moons vary from being fully resolved disks to point sources. Due to pointing, timing, and other errors, the moons are not always exactly where SPICE predicts them to be. To refine their position, we implement a number of techniques to search for the moon within a box around the SPICE predicted position. The box is the larger of a 20$\times$20~pixel box or a box that is twice the anticipated radius of the moon. To start, we use a broad (10~pixel) Gaussian smoothing filter to identify and remove hot pixel sources, such as bright cosmic rays and stars, and then determine the location of the brightest pixel in the box, which we assume is the moon.

\begin{figure*}
\plottwo{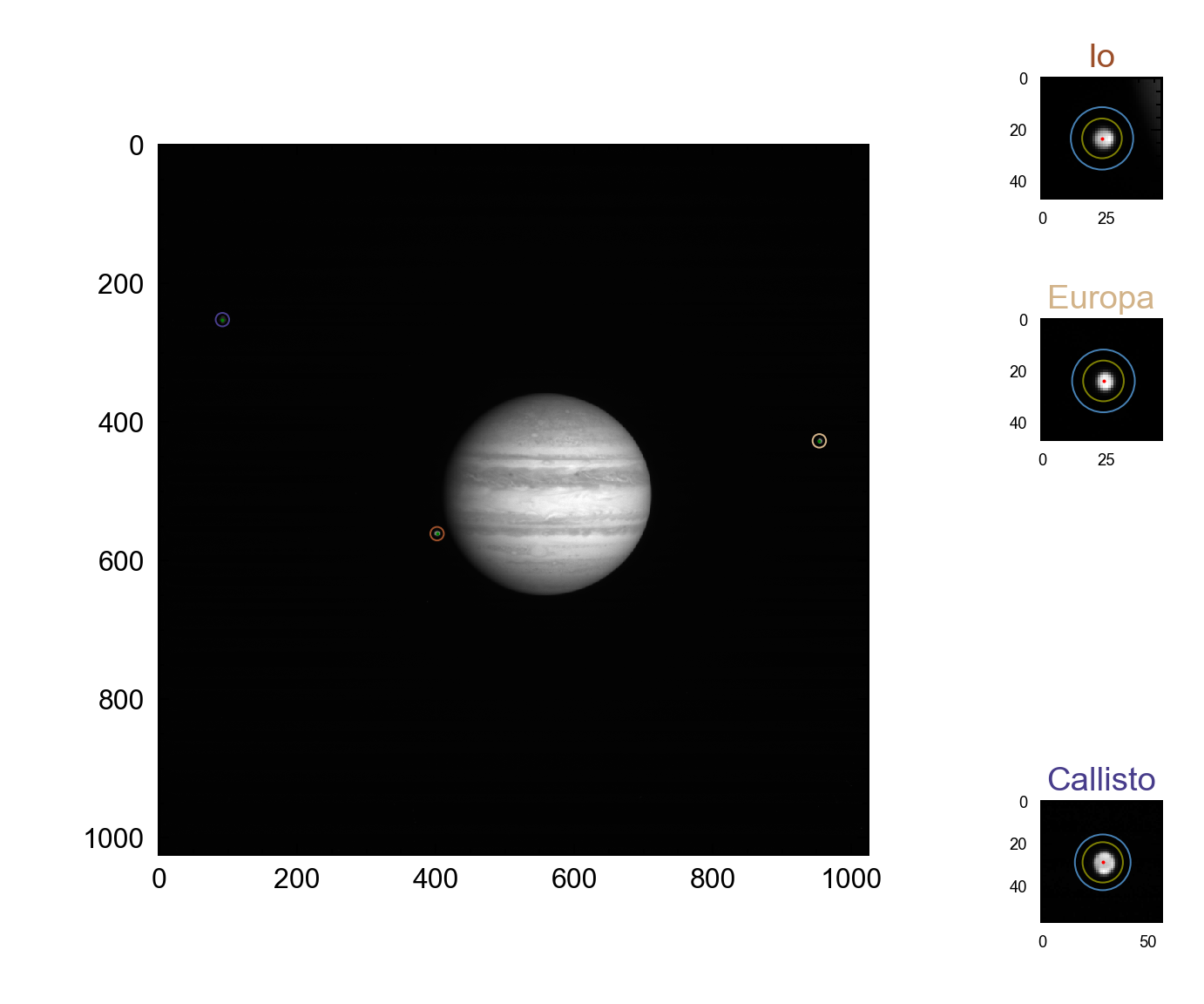}{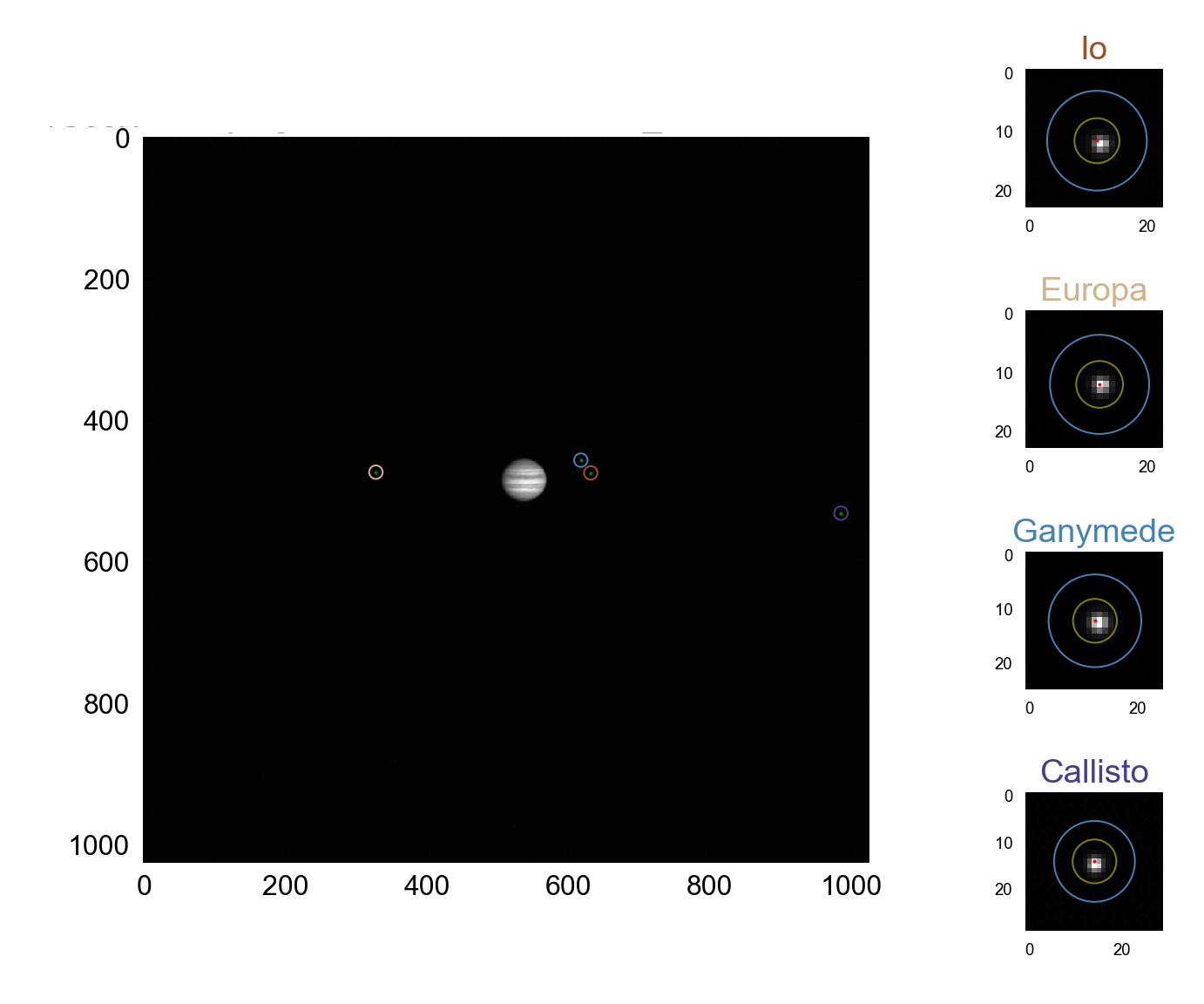}
\caption{Examples of reduction pipeline diagnostic images. Left: A sample NAC image of Jupiter. The various circles show the expected positions of the moons. The postage stamps show the approximate center (red point) of the aperture (green) and inner sky annulus (blue). Ganymede was not found/present in this image. Right: same as left but for a WAC image. \label{fig:hough}}
\end{figure*}

In instances where the moon is sufficiently distant as to be unresolved in the image or less than 5~pixels, we simply set the maximum within the box as the location of the moon. For instances where the moon is expected to be larger than 5~pixels, we employ the circle-finding algorithms in the \verb|scikit-image| Hough transformation routines, a set of feature extraction techniques used commonly in computer vision work, which result in a position and radius of a circle around the moon and is fairly robust to moon phases. An example of the algorithm at work at a high phase angle is shown in \autoref{fig:hough}.

If after either center finding algorithm the moon is within 5~pixels of the edge of the image, we discard the image. 

\verb|CISSCALv3.9| computes the reflectivity, $\frac{I}{F}$, the intensity of the detected light, $I$, normalized to the incident solar flux, $F$. The reduction pipeline \citep{Knowles2016} assumes a solar spectrum and the distance to Jupiter. Thus, we rescale $F$ to each moon's distance from the Sun and the spacecraft before before conducting aperture photometry. Using an HST spectrum of Enceladus \citet{Knowles2016} determined the filter-appropriate radiometric correction factors to ensure this $\frac{I}{F}$ is a measure of albedo.\deleted{ Therefore, it should be noted that these data are not absolute flux corrected.} \citet{Mayorga2016} presented normalized phase curves (phase functions) that were then scaled to the reference spectra from \citet{Karkoschka1994, Karkoschka1998}, because these radiometric correction factors were not included in \verb|CISSCALv3.6|. Here, we present these data without normalization and no scaling to any reference spectrum using only radiometric correction included with \verb|CISSCALv3.9|.

Since the point spread function of \CISS{} is not well known (see discussion in \cite{Mayorga2016}), to determine the minimum aperture size, we took images of all the moons at phase angles less than 20$\degr{}$ and tested various aperture sizes, $a_{\rm p}$. We generated curves of measured reflectivity for apertures ranging from 1--10~pixels in 0.5~pixel increments. With the ensemble of images we determine the mean base aperture size that would result in the inclusion of 95\% of the light, a reasonable compromise to avoid the inclusion of background noise that actually includes all of the lunar light when viewed by eye (see \autoref{fig:hough}). The minimum radius of the aperture that would include 99.7\% of the light is the beginning of the base sky annulus, $a_{\rm sky}$. The outer sky annulus is 3~pixels larger. We choose to keep the sky annulus close and small to allow for photometry of moons that appear very close together, ensuring less contamination from background stars, and minimizing the banded background variations \citep[see the 2~Hz noise in][]{Knowles2016}. Note that these base radii are then increased depending on the properties of the moons. We use the \verb|photutils.aperture_photometry| routine and other ancillary routines to perform aperture photometry.

Depending on the unresolved or resolved nature of the moon, we then increase the minimum aperture and sky annulus radius appropriately. For unresolved moons, i.e. the moon's radius is less than twice the full-width-half-max of the point spread function of the filter in question, and for resolved moons less than 5~pixels, we increase the aperture by one-fifth the radius of the moon. For resolved moons larger than 5~pixels, we use an aperture that is 1.05 times larger than the circle returned by the Hough circle-finding algorithm and a sky annulus 1.1 times larger. The aperture sizes for each moon are shown in \autoref{tbl:aps}. We eliminate images from the sample where the sky background computed was more than 50\% of the measured flux from the moon, typically the moons are underexposed in these images or there is contamination from Jupiter in the foreground or background. This is the case for nearly half of the images because Jupiter was the target of the majority of the observations.

\begin{deluxetable}{lrrrrrr}
\tabletypesize{\scriptsize}
\tablecaption{Aperture and sky annulus radii, where $R_m$ is in pixels. \label{tbl:aps}}
\tablehead{\colhead{Moon} & \multicolumn{2}{c}{Base} & \multicolumn{2}{c}{$<$5 pix} & \multicolumn{2}{c}{$>$5 pix}\\
& \colhead{$a_{\rm p}$} & \colhead{$a_{\rm sky}$} & \colhead{$a_{\rm p}$} & \colhead{$a_{\rm sky}$} & \colhead{$a_{\rm p}$} & \colhead{$a_{\rm sky}$}}
\startdata
Io & 3.543 & 7.767 & \multirow{4}{*}{0.2R$_m$} & \multirow{4}{*}{0.5R$_m$} & \multirow{4}{*}{1.05R$_m$} & \multirow{4}{*}{1.1R$_m$}\\
Europa & 3.733 & 7.820 & & & &\\
Ganymede & 3.646 & 7.549 & & & &\\
Callisto & 4.293 & 7.672 & & & &\\
\enddata
\end{deluxetable}

As a final step, we visually inspect the resulting 5,487 images and remove any cases where: the predicted moon size is discrepant with its appearance in the image, which is a symptom of improper identification; where the moon centering failed, which occurs commonly on high phase angle images where the night side is difficult for the algorithm to distinguish the moon from the background or due to position errors; and where there is clear background or foreground contamination from Jupiter, background stars, oversaturated pixels, and/or transient activity (such as cosmic rays) that was not caught by our automate procedures. This results in 4,847 images remaining in the dataset. 

\begin{figure}
\plotone{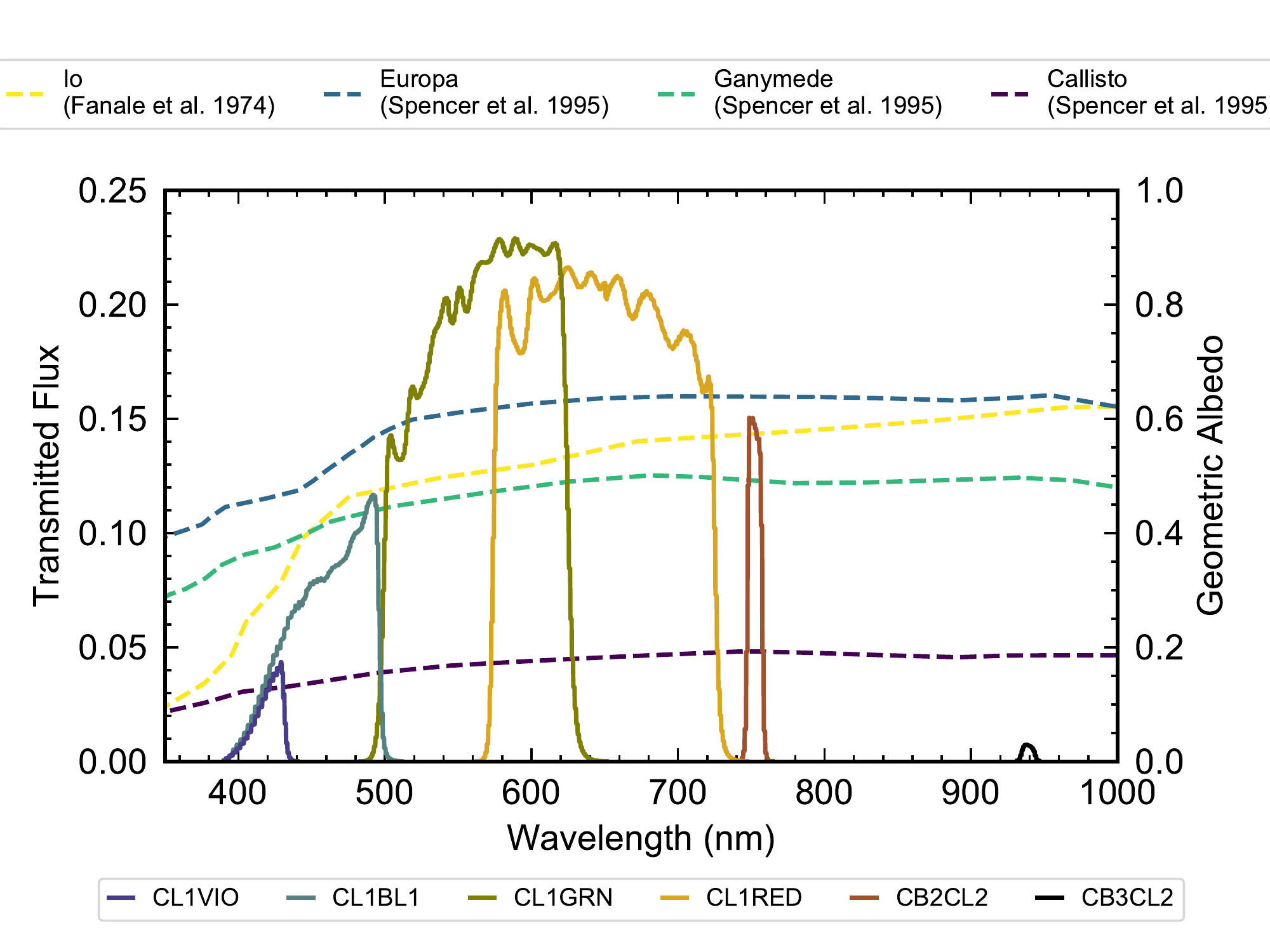}
\caption{The WAC filter transmission curves that are included in this study. Note: the spectral transmission between the two cameras are slightly different \citep[see][]{Porco2004}. We also plot the albedo spectra of the Galilean satellites for comparison.\label{fig:trans}}
\end{figure}

The most common filters in the final image set are VIO\footnote{VIO exists only on the WAC.}, BL1\footnote{The blue filter is on a different wheel in either camera such that the combination is CL1BL1 and BL1CL2. Additionally, the effective wavelength of the filter in the NAC is the most different from the WAC for all the filters we consider here (8~nm).}, GRN, RED, CB2, and CB3 in combination with a clear filter. We use these six filters in our study to preserve a wide phase angle and longitudinal dataset. The majority of the images are from the WAC (3,299) with only a few imaged from the NAC (329). The NAC images are predominantly taken with BL1 or GRN. Again, due to the slightly differing optical setups the transmission functions are slightly different. The WAC versions of the transmission curves for these filters are shown in \autoref{fig:trans}. We also plot a set of albedo spectra for the Galilean satellites for comparison as compiled in \citet{Madden2018} (ultimately, \citet{Fanale1974} for Io and \citet{Spencer1995} for the other moons). Note that the BL1 filter includes the same wavelengths as the VIO filter with a very similar transmission function in that wavelength range. The dataset by moon and filter combination are quantified in \autoref{tbl:count}. The effective wavelengths were computed using the full system transmission function convolved with a solar spectrum \citep{Porco2004}.

\begin{deluxetable*}{lCCCCCCR}
\tablecaption{Number of observations in a filter for a given moon and the filter's effective wavelength in nm. \label{tbl:count}}
\tablehead{\colhead{Moon} & \multicolumn{6}{c}{Filter} & \colhead{Total} \\ 
& \mathrm{VIO} & \mathrm{BL1} & \mathrm{GRN} & \mathrm{RED} & \mathrm{CB2} & \mathrm{CB3} & \\
\cline{2-7}
& 420 & 455\tablenotemark{a},463 & 569\tablenotemark{a},568 & 647 & 752 & 939 & {\rm[nm]}}
\startdata
Io & 585 & 85,85 & 64,610 & 565 & 328 & 188 & 2510 \\
Europa & 622 & 45,60 & 36,631 & 599 & 343 & 130 & 2466 \\
Ganymede & 574 & 62,62 & 47,588 & 564 & 293 & 202 & 2392 \\
Callisto & 396 & 18,1 & 14,379 & 373 & 123 & 3 & 1307 \\
\hline WAC Images & 809 & 145 & 833 & 764 & 448 & 300 & 3299 \\
NAC Images & 0 & 187 & 142 & 0 & 0 & 0 & 329 \\
\enddata
\tablenotetext{\rm a}{These NAC effective wavelengths differ slightly from the WAC.}
\end{deluxetable*}

The moons are viewed from approximately the equator, the point on the moon directly below the spacecraft has a latitude smaller than 4$\degr{}$ north or south throughout the whole flyby. The majority of the phase angle coverage is from 0$\degr{}$--25$\degr{}$. There are some observations from 60$\degr{}$--135$\degr{}$ that can be used to constrain the phase functions of each of the moons but there is a notable absence of data between about 30$\degr{}$--60$\degr{}$ where Jupiter would have filled the entire field of view. A subset of the data are tabulated in \autoref{tbl:phot}.

\begin{deluxetable*}{lcj{3.3}j{3.3}j{3.3}r}
\tablecaption{Tabulated reflectivity as a function of moon, filter, phase angle, and  planetocentric longitude.\label{tbl:phot}}
\tablehead{\colhead{Moon} & \colhead{Filter} & \colhead{Phase Angle ($\degr{}$)} & \colhead{Planetocentric Longitude ($\degr{}$)} & \colhead{Reflectivity ($\frac{I}{F}$)} & \colhead{Image}}
\startdata
Io & CL1VIO & 20.462\cdots & 65.961\cdots & 0.148\cdots & W1349093639\_2.IMG \\
Europa & CL1GRN & 20.626\cdots & 108.193\cdots & 0.458\cdots & W1349111745\_2.IMG \\
Ganymede & CL1VIO & 14.833\cdots & 156.352\cdots & 0.250\cdots & W1353910514\_1.IMG \\
Ganymede & CB2CL2 & 3.516\cdots & -147.025\cdots & 0.435\cdots & W1355071267\_1.IMG \\
Callisto & CL1RED & 16.531\cdots & 169.563\cdots & 0.120\cdots & W1353461870\_1.IMG \\
\enddata
\tablecomments{\autoref{tbl:phot} is published in its 8000 line entirety in machine-readable format at machine precision. A random set of rows is shown here for guidance regarding its form and content.}
\end{deluxetable*}

\section{Modeling}
\label{sec:model}
With sufficient monitoring of a planet, it is possible to measure the rotational light curves of non-tidally locked planets using high precision stellar light curves or direct imaging. Although the Galilean satellites are tidally locked with a small eccentricity, Jupiter is not the source of light. Thus, we can disentangle reflectivity variations caused by illumination as a function of phase angle from surface variations as a function of planetocentric longitude. Since we have multiple filters available, we can also look at the color variations with phase angle and planetocentric longitude. The light curves measured by \CISS{} are a combination of reflectivity variations with rotational variations superimposed with the illumination variations (i.e. the phase curve). To disentangle them, our procedure is as follows:

\begin{enumerate}
\item Fit the illumination variations with an initial third order polynomial phase curve, $f(\alpha)$, across the entire phase angle range.
\item \label{itm:phasefit} Divide the phase curve fit from the phase angle range of \parange{} \added{to remove the effect of illumination variations in preparation to fit the rotational variations}.
\item Fit the rotational variations with the rotational brightness model explained in \autoref{ssec:rot-model}.
\item Remove the rotational variations from the illumination variations where $\alpha<25\degr{}$.
\item Fit the phase curve across the entire range of available phase angles using polynomials of 1st-10th order and select the optimal fit based on the Bayes Information Criterion (BIC).
\item Repeat from \autoref{itm:phasefit}.
\end{enumerate}

When fitting polynomials we restrict the phase curve fit to be 0 at a phase angle of 180$\degr{}$ ($f(180\degr{}$)=0). We assume that our errors are normally distributed and compute the Bayesian Information Criterion (BIC),
\begin{equation}
    BIC = N \ln{\left(
        \frac{1}{N} \sum_{i=1}^N \left(\frac{I_i}{F} - {f(\alpha_i)}\right)^2
    \right)} + k\ln{N},
\end{equation}
for each polynomial, where $N$ is the number of data points, $f(\alpha)$ is the resulting fit on the data, and $k$ is the polynomial's order. Typically, the BIC decreases monotonically before plateauing with only small decreases or increases thereafter. We select as our best fit the order where the slope falls to less than 2\% the overall change in BIC.

Since the majority of the data covers phase angles from 0\degr{}-25\degr{}, we consider rotational variations at only these angles. We narrow our rotational fit range to \parange{} to minimize amplitudinal variations caused by illumination as well as maximize the data used in the fit. It should be noted that in this phase angle range, \citep{Mayorga2016} observed a deviation and an increase in the scatter of reflectivity measurements of Jupiter in the RED and GRN filters that is caused by inconsistencies in the camera shutter speed \citep{Mayorga2016}. This occurs in images of 5~ms duration that were taken on approach of Jupiter and is similarly observed for the moon data.

To treat all moons equally, this iterative process was only conducted for moon and filter combinations that had sufficient phase angle and longitudinal coverage. Notably, while there may be sufficient data for BL1 phase curves of all moons, there is insufficient longitudinal coverage for any of them. Phase angle coverage in the CB2 and CB3 filters is also limited to roughly $\alpha<25\degr{}$, but longitudinal coverage is good for some of the moons and thus we proceed in the iterative process when possible.

We fit the rotational light curves with a simple orange slice model following the work of \citet{Cowan2008} as described in \autoref{ssec:rot-model}. We fit for the brightness of N longitudinal slices for each of the moons available filters using \verb|PlanetSlicer| \citep{PlanetSlicer}. The resulting map we infer from the data we term the \emph{inverted map} following \citet{Cowan2008}. When we compared the $\chi^2$ achieved from varying the number of slices, we find that the Io, Europa, Ganymede, and Callisto solutions do not offer significant $\chi^2$ improvements beyond \slices{} slices, respectively, so we opt for N=6. Other mapping techniques have been proposed since the onset of this work \citep[c.f.][]{STARRY, Luger2019, Fan2019} that are promising to address this complex problem for exoplanets.

\subsection{Rotational Brightness Model}
\label{ssec:rot-model}
Expanding on the model of \cite{Cowan2008}, we construct a model for the brightness of the object given the planetocentric longitude of insolation $\phi_\mathrm{in}$, the planetocentric longitude corresponding to the observer $\phi_\mathrm{obs}$, and the albedo of the planet, $A_i$, the proportion of the incident light or radiation that is reflected by a surface, which is divided into a set of longitudinal slices $i$.  Assuming Lambertian scattering, the brightness contribution from a single point on the planet is the product of the insolation, albedo, and area element (from the observer's perspective) at that point. \added{We will show later that the Galilean moons are non-Lambertian, but make the choice here for the simplicity of its analytic solution.} The brightness of the surface at a given point is the incoming flux, $F$, which is the incident flux from the star $F_*$ multiplied by the surface area factor $\cos(\phi-\phi_\mathrm{in})\sin(\theta)$, where $\theta$ is the latitude, modulated by the albedo.  The area element for the observer is similarly $\cos(\phi-\phi_\mathrm{obs})\sin(\theta)$.  We can integrate this to obtain the total contribution from a single slice to the observed brightness $J_i$.

\begin{align}
    J_i &= \int_{\phi_-}^{\phi_+} \int_0^\pi
        A_i F(\phi,\phi_\mathrm{in})
        \cos(\phi-\phi_\mathrm{obs})
        \sin^2(\theta)
    d\theta d\phi\\
    &= \frac{4 F_* A_i}{3}\int_{\phi_-}^{\phi_+}
        \cos(\phi-\phi_\mathrm{in})
        \cos(\phi-\phi_\mathrm{obs})
    d\phi\\
    &= \left. \frac{F_* A_i}{3} \left(
        2 \phi \cos(\phi_\mathrm{in} - \phi_\mathrm{obs})
        - \sin(\phi_\mathrm{obs} + \phi_\mathrm{in} - 2\phi)
    \right) \right|_{\phi_-}^{\phi_+}\label{eq:modelCoefficients}
\end{align}

The longitudinal bounds of the integral $\phi_+$ and $\phi_-$ are the edges of the slice clipped to exclude regions that are not illuminated or not visible to the observer \citep[similar to][]{Cowan2008}.  For example, if an entire slice is on the far side of the planet from the observer, $\phi_+ = \phi_-$ and $J_i = 0$.

Note that $J_i$ is linear with $A_i$.  Thus we can construct a matrix $\mathit{G}$ from \autoref{eq:modelCoefficients} whose rows are longitudinal slices and whose columns are observations, such that $\mathbf{J} = \mathit{G} \mathbf{A}$.  Here $\mathbf{J}$ the total brightness for each observation and $\mathbf{A} = [A_0, A_1, A_2, ...]$ is the albedo for each slice.  Because $0 \leq A_i \leq 1$, we solve for $\mathbf{A}$ from $\mathbf{J}$ using the bounded least squares solver \verb|lsq_linear| from Scipy \citep{SciPy}.  This type of model can exhibit degeneracies in $\mathbf{A}$ \citep[see][]{Knutson2007}, usually necessitating a prior on the components of $\mathbf{A}$.  However, unlike the hot Jupiters considered in \cite{Cowan2008} and \cite{Knutson2007}, $\phi_\mathrm{in}$ is not always constant.  As the phase angle $\alpha = \phi_\mathrm{in} - \phi_\mathrm{obs}$ varies, the contribution of each slice to the total brightness changes, modifying $\mathit{G}$ and breaking the model degeneracy.  Thus, it is in general easier to observe objects that are not tidally locked to their parent stars (such as moons). With this model we can \emph{invert} the light curve to achieve a brightness map of the satellites.

\section{Results}
\label{sec:Results}

\begin{figure*}
\plotone{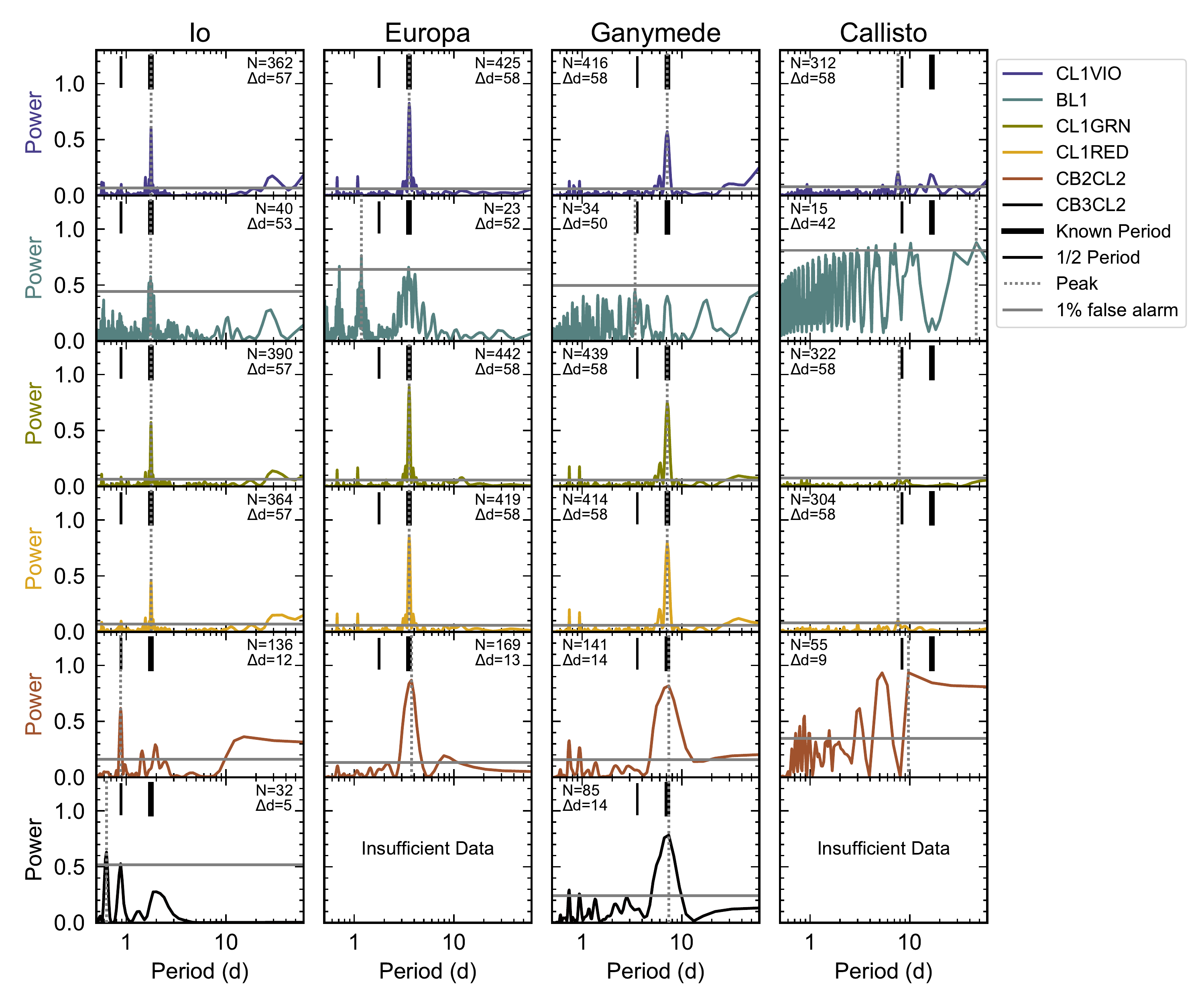}
\caption{The Lomb-Scargle periodograms for each moon from top to bottom in the filters of VIO, BL1, GRN, RED, CB2, and CB3. The black solid dashes indicate the known period and half the known period. The gray dotted line marks the measured peak in the periodogram. The gray horizontal line marks the 1\% false alarm probability. The rotation periods are well recovered without a phase angle correction in VIO, GRN, RED, with the exception of Callisto. Recovering the rotation period of Callisto (and in other filters) is stymied by the observational baseline being too short and/or poor sampling.\label{fig:lomb}}
\end{figure*}

All the moons demonstrate variations as a function of planetocentric longitude which allow their rotation periods to be determined \citep[c.f.][]{Fujii2014}. A Lomb-Scargle periodogram  readily reproduces the rotation periods, even without applying a phase angle correction, which is promising for the detection of rotation periods of directly imaged exoplants. The periodograms for each of the moons are shown in \autoref{fig:lomb}, and we compute the percent disagreement between the true period and the detected result. We find the disagreement below the percent level for all moons but Callisto in the VIO, GRN, and RED filters, likely because of the smaller amplitude of the variations compared to the scatter of the observations and the fewest number of rotation periods observed. The BL1, CB2, and CB3 results are likely due to a poor sampling rate across the observational time baseline as well as having a very short observing baseline. Applying the phase angle correction strengthens the detection of the rotation periods of all the moons including Callisto. 

We analyze the reflectivity variations as a function of rotation at low phase angles (\autoref{ssec:rotational}), the reflectivity variations as a function of illumination and compare to a Lambertian phase curve (\autoref{ssec:phasecurve}), the reflectivity variations as a function of rotation at high phase angles (\autoref{ssec:highphase}), and color variations as a function of illumination and rotation (\autoref{ssec:color}).

\subsection{Rotational Variations}
\label{ssec:rotational}
\begin{figure*}
\plotone{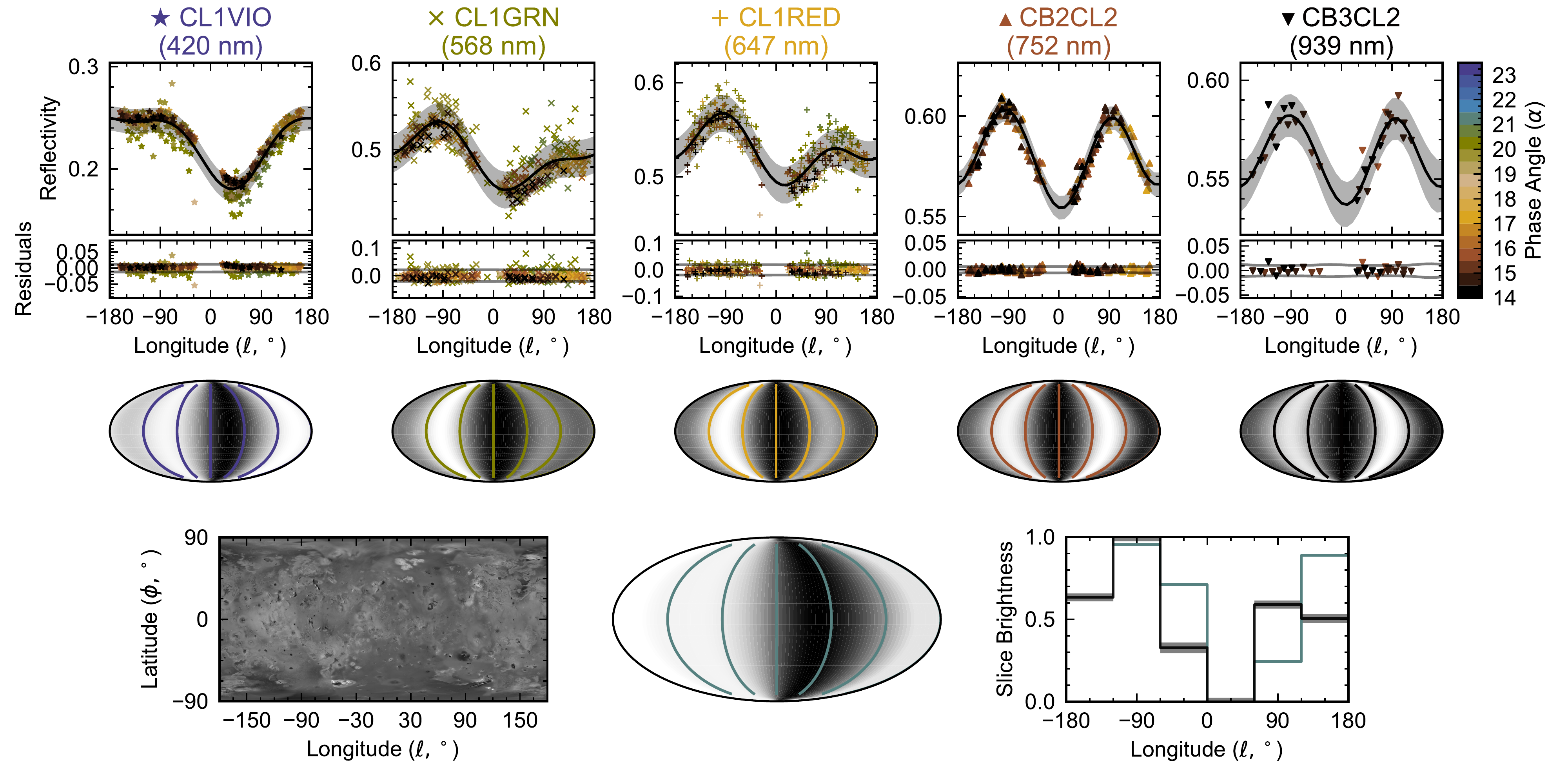}
\caption{Variation of the brightness of Io as a function of longitude. Top: the reflectivity as a function of planetocentric longitude \added{of the sub-spacecraft point} of Io (points) and colored by phase angle. The resulting fit is shown in black for (from left to right) the VIO, GRN, RED, CB2, and CB3 filters. The shaded region marks the 1$\sigma$ error region. Center: the inverted brightness maps for each filter. The brightness slices have a been interpolated between each defining longitude to smooth the appearance of the map and colored lines indicate the edges of the slices. Bottom: the USGS Map (left), the resulting sliced, interpolated, and smoothed map (center), and the slice brightness comparison where the USGS map result is in blue and the linear combination of the \emph{inverted map} results is shown in black with 1$\sigma$ errors shaded. For the comparison, the slice brightnesses were normalized to peak at 1.\label{fig:io}}
\end{figure*}

\begin{figure*}
\plotone{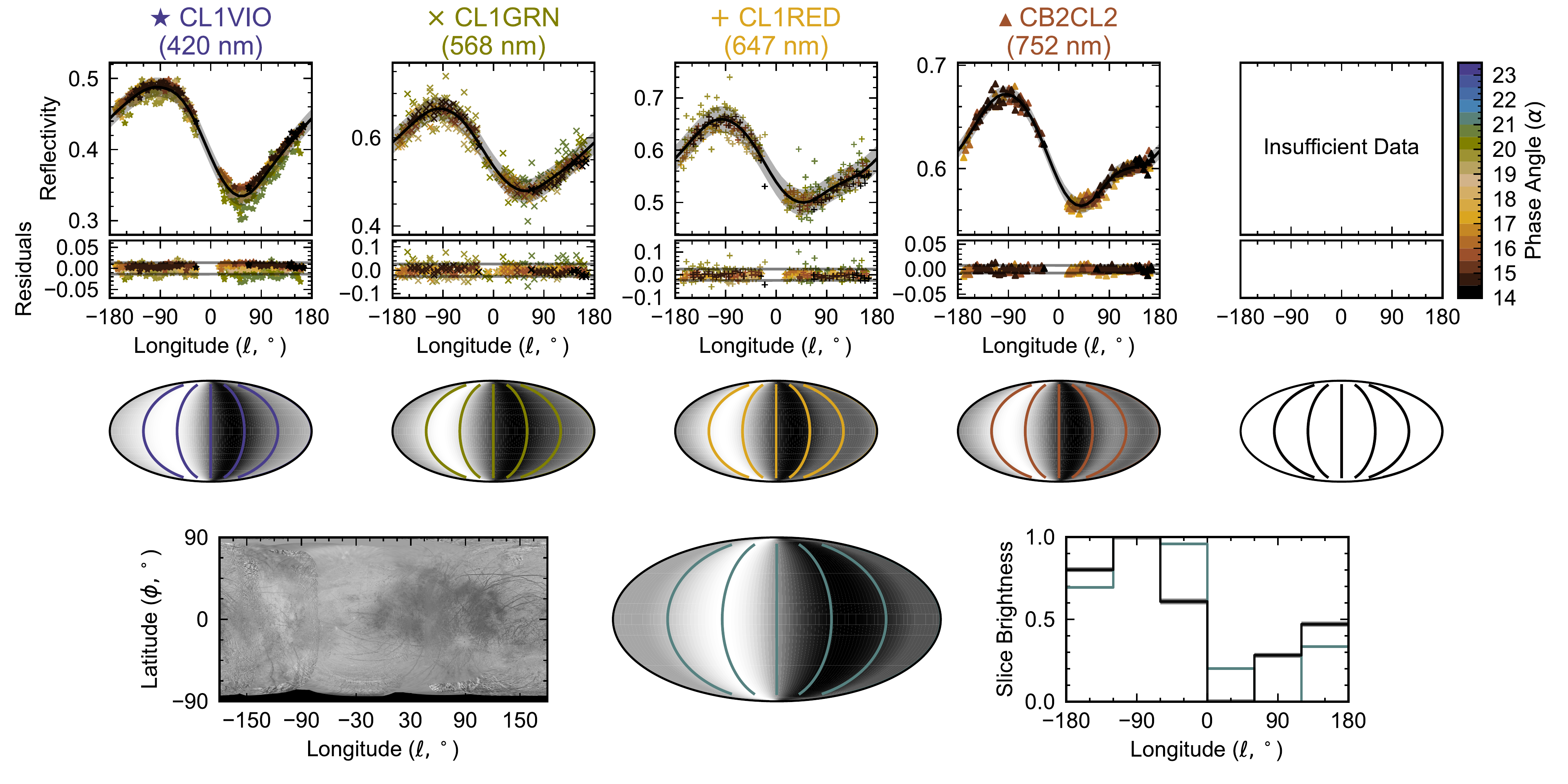}
\caption{Variation of the brightness of Europa as a function of planetocentric longitude; see the description in the caption of \autoref{fig:io}.\label{fig:europa}}
\end{figure*}

\begin{figure*}
\plotone{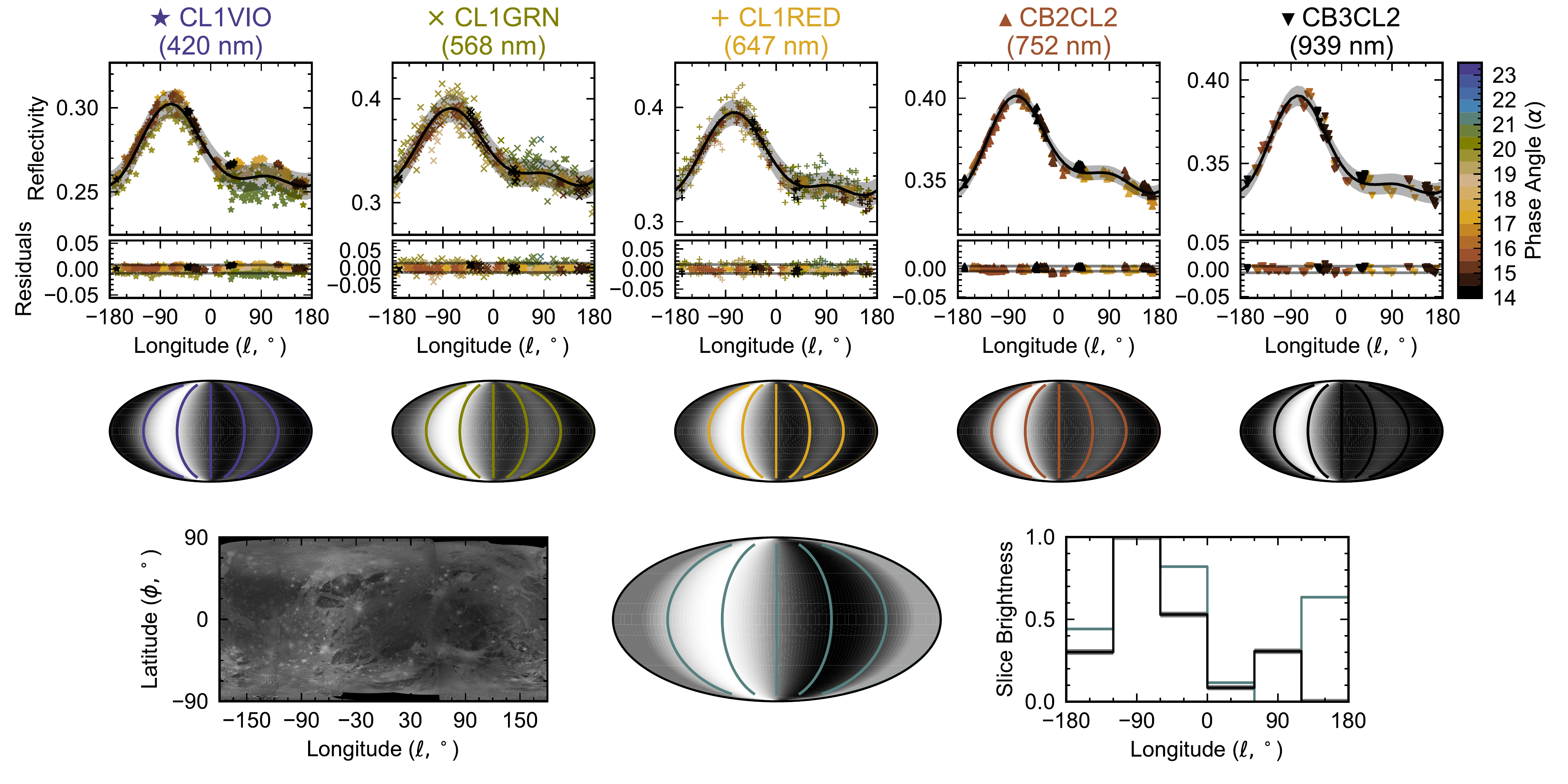}
\caption{Variation of the brightness of Ganymede as a function of planetocentric longitude; see the description in the caption of \autoref{fig:io}.\label{fig:ganymede}}
\end{figure*}

\begin{figure*}
\plotone{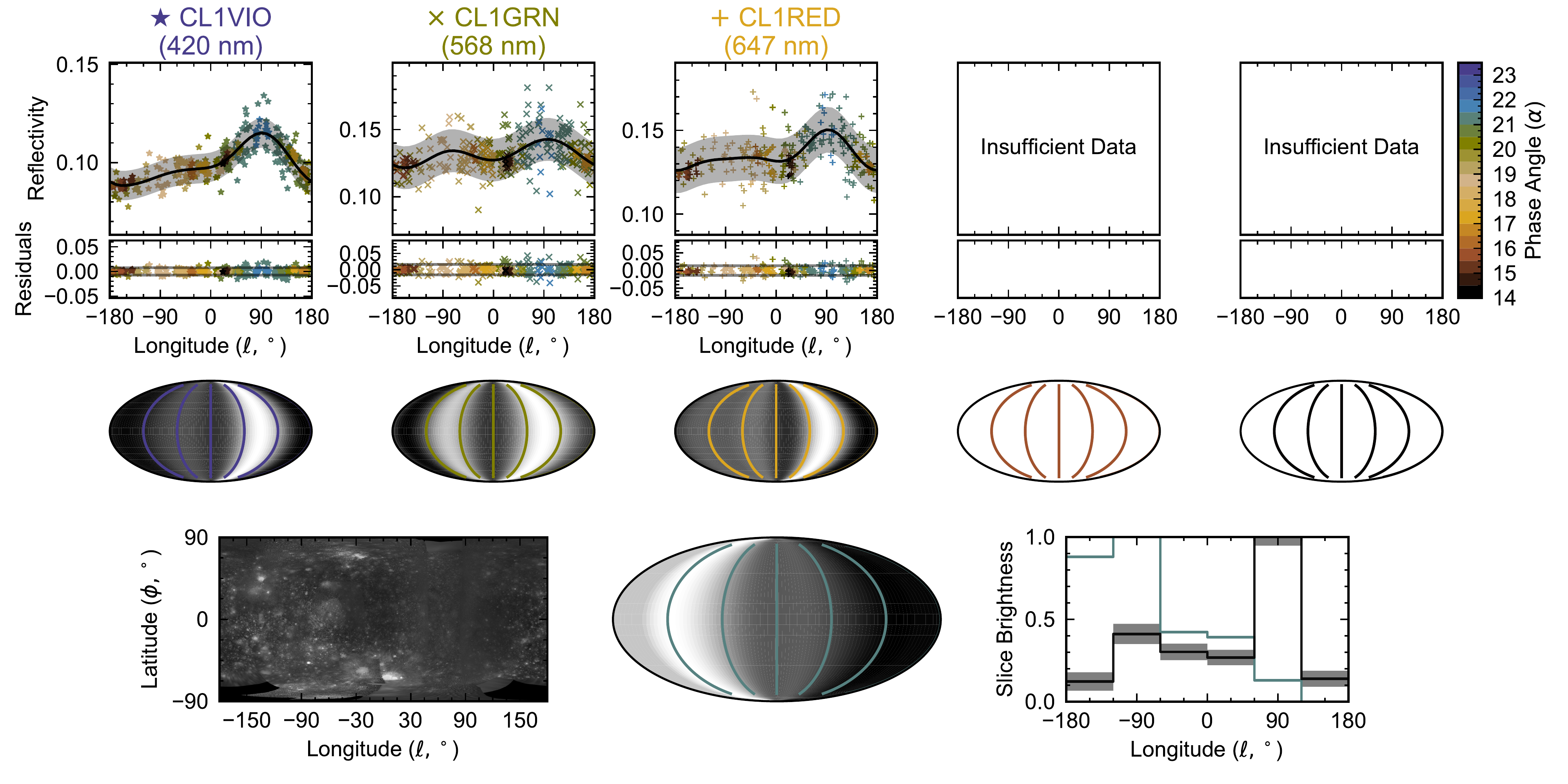}
\caption{Variation of the brightness of Callisto as a function of planetocentric longitude; see the description in the caption of \autoref{fig:io}. Note that the amplitude of the variations are comparable to the error.\label{fig:callisto}}
\end{figure*}

While variations would be proportionally larger when viewing a planet at high phase angles (crescent), the majority of the phase angle coverage is between 0$\degr{}$--25$\degr{}$. These data span approximately 60 days between October to late November 2000. The rotational data and resulting fit for each moon using \verb|PlanetSlicer| are shown in the top panels of Figures \ref{fig:io}--\ref{fig:callisto}. The resulting brightness maps for each moon are shown in the middle panels with a smoothing across planetocentric longitude.

For visual comparison, we use the mosaics produced by the US Geological Survey \citep[USGS, e.g.][]{Williams2011, EuropaMap, GanymedeMap, CallistoMap}. The mosaics are composed of Voyager and Galileo data typically taken in a clear filter with other filters to supplement gaps or where doing so would increase the resolution of the mosaic. The images were empirically adjusted in brightness and contrast to match in overlapping areas \citep[see][]{Mosaics}. Note that the planetocentric longitude system conventions differ for each mosaic and we have translated them into the E-W system (-180$\degr{}$ to 180$\degr{}$).

The USGS mosaics are not albedo measurements, they are grayscale color values in a \verb|.tif|. Also, since the images composing the mosaics have various illumination angles, resolutions, and were taken in varying filters, these maps cannot be compared directly with our inverted brightness maps \citep[see][]{Fujii2014}. As such, the slice brightnesses are only correct relative to each other, but inherently meaningless as an albedo measurement. Green filter images were most often used to replace clear images but other \emph{Voyager} and \emph{Galileo} filters were used as well to increase the spatial resolution of the map. We use a common color theory practice for image file types to determine the average brightness of a slice, $\overline{B_j}$
\begin{equation}
\overline{B_j} = \sqrt{\sum_i \frac{p_{j,i}^2}{N}}
\end{equation}, where $p_{j,i}^2$ is the value of the $i$th pixel in slice $j$ \citep[see][for more on gamma corrections]{McREYNOLDS2005}. We also apply the appropriate area projection correction with latitude to reflect an equatorial viewing perspective and weigh polar information less heavily. The USGS map is shown in the bottom left panel in Figures \ref{fig:io}--\ref{fig:callisto} with the integrated version in the bottom center panel. A comparison in the relative slice brightnesses of the mosaic and a linear combination of the fits from the VIO, GRN, and RED filters (VIO+GRN+RED) is shown the bottom right panel. It should be noted that \emph{Cassini} is more sensitive to red wavelengths than \emph{Voyager}. To qualitatively compare with the slice brightnesses from the inverted brightness maps, we scale the slice brightnesses from the USGS map to unity.

For Io, the shape and amplitude of the \replaced{phase curve}{rotational variations} can be seen to vary with wavelength and the largest amplitude is in the VIO filter. From VIO to GRN the shape of the \added{rotational} phase curve changes from essentially a single peaked feature to a double peaked shape. The inverted brightness maps demonstrate a minimum where also predicted by the USGS map.

Europa, being icy, has a very sharp variation in its \replaced{phase curve}{rotational variations} suggestive of strong backscattering from planetocentric longitudes with snowy or icy plains. The largest variation is in the VIO filter. The inverted brightness maps demonstrate a minimum where also predicted by the USGS map, but have a slightly shifted planetocentric longitude marking the peak albedo.

For Ganymede, the shape of the \replaced{phase curve}{rotational variations} is fairly consistent with wavelength. The amplitudes are also very consistent between wavelengths and the VIO filter has the smallest amplitude. The inverted brightness maps finds a similar longitude for the maximum but has an absolute minimum near 180\degr{} and only a local near 0\degr{}.

For Callisto, the shape of the \replaced{phase curve}{rotational variations} is reminiscent of a doubly peaked sinusoid, but the scatter is sufficient to limit our interpretation of the amplitude and wavelength variations. The VIO filter (and the others less confidently) suggests that the brightest hemisphere is the eastern hemisphere, exactly opposite the USGS map. The USGS map for Callisto is provided in the same longitudinal coordinate system as the Europa and Ganymede maps, and in fact, Io is the map provided in a flipped coordinate system. We can only attribute this discrepancy to the scatter in the data combined with our underlying Lambertian phase function assumptions for the orange slice model.

We provide the brightness slice values for \verb|PlanetSlicer| in \autoref{tbl:slicer} so that they can be used to reproduce our fits for the variations with rotation.

\begin{deluxetable*}{lcLLLLLL}
\tablecaption{The slice brightness values derived by \texttt{PlanetSlicer}. \label{tbl:slicer}}
\startdata
\tablehead{\colhead{Moon} & \colhead{Filter} & \colhead{$J_0$} & \colhead{$J_1$} & \colhead{$J_2$} & \colhead{$J_3$} & \colhead{$J_4$} & \colhead{$J_5$}}
\multirow{5}{*}{Io} & CL1VIO & 0.117327 & 0.126080 & 0.109350 & 0.076307 & 0.107476 & 0.127953 \\
 & CL1GRN & 0.250041 & 0.273365 & 0.228565 & 0.215799 & 0.244837 & 0.236488 \\
 & CL1RED & 0.264435 & 0.292585 & 0.243468 & 0.235477 & 0.272270 & 0.246266 \\
 & CB2CL2 & 0.285379 & 0.306900 & 0.268862 & 0.271947 & 0.309278 & 0.269084 \\
 & CB3CL2 & 0.275214 & 0.295680 & 0.263344 & 0.260699 & 0.301027 & 0.259521 \\
\multirow{4}{*}{Europa} & CL1VIO & 0.232966 & 0.246369 & 0.221652 & 0.146874 & 0.179195 & 0.209323 \\
 & CL1GRN & 0.313415 & 0.341145 & 0.287923 & 0.222243 & 0.245618 & 0.273885 \\
 & CL1RED & 0.312869 & 0.336263 & 0.286555 & 0.228525 & 0.264607 & 0.268065 \\
 & CB2CL2 & 0.320300 & 0.339759 & 0.305061 & 0.260918 & 0.301074 & 0.289561 \\
\multirow{5}{*}{Ganymede} & CL1VIO & 0.126857 & 0.157164 & 0.139248 & 0.122491 & 0.130721 & 0.121970 \\
 & CL1GRN & 0.168968 & 0.203035 & 0.178112 & 0.155621 & 0.166954 & 0.147735 \\
 & CL1RED & 0.169613 & 0.205668 & 0.180834 & 0.156233 & 0.168460 & 0.152404 \\
 & CB2CL2 & 0.172505 & 0.209228 & 0.182066 & 0.170910 & 0.177491 & 0.163869 \\
 & CB3CL2 & 0.166920 & 0.204552 & 0.175288 & 0.162750 & 0.168631 & 0.160370 \\
\multirow{3}{*}{Callisto} & CL1VIO & 0.042274 & 0.046918 & 0.047588 & 0.048241 & 0.062808 & 0.044274 \\
 & CL1GRN & 0.056125 & 0.070270 & 0.060978 & 0.063225 & 0.074481 & 0.063077 \\
 & CL1RED & 0.065258 & 0.065042 & 0.066643 & 0.061564 & 0.082743 & 0.057380 \\
\enddata
\end{deluxetable*}

\subsection{Phase Curves}
\label{ssec:phasecurve}
\begin{figure*}
\plotone{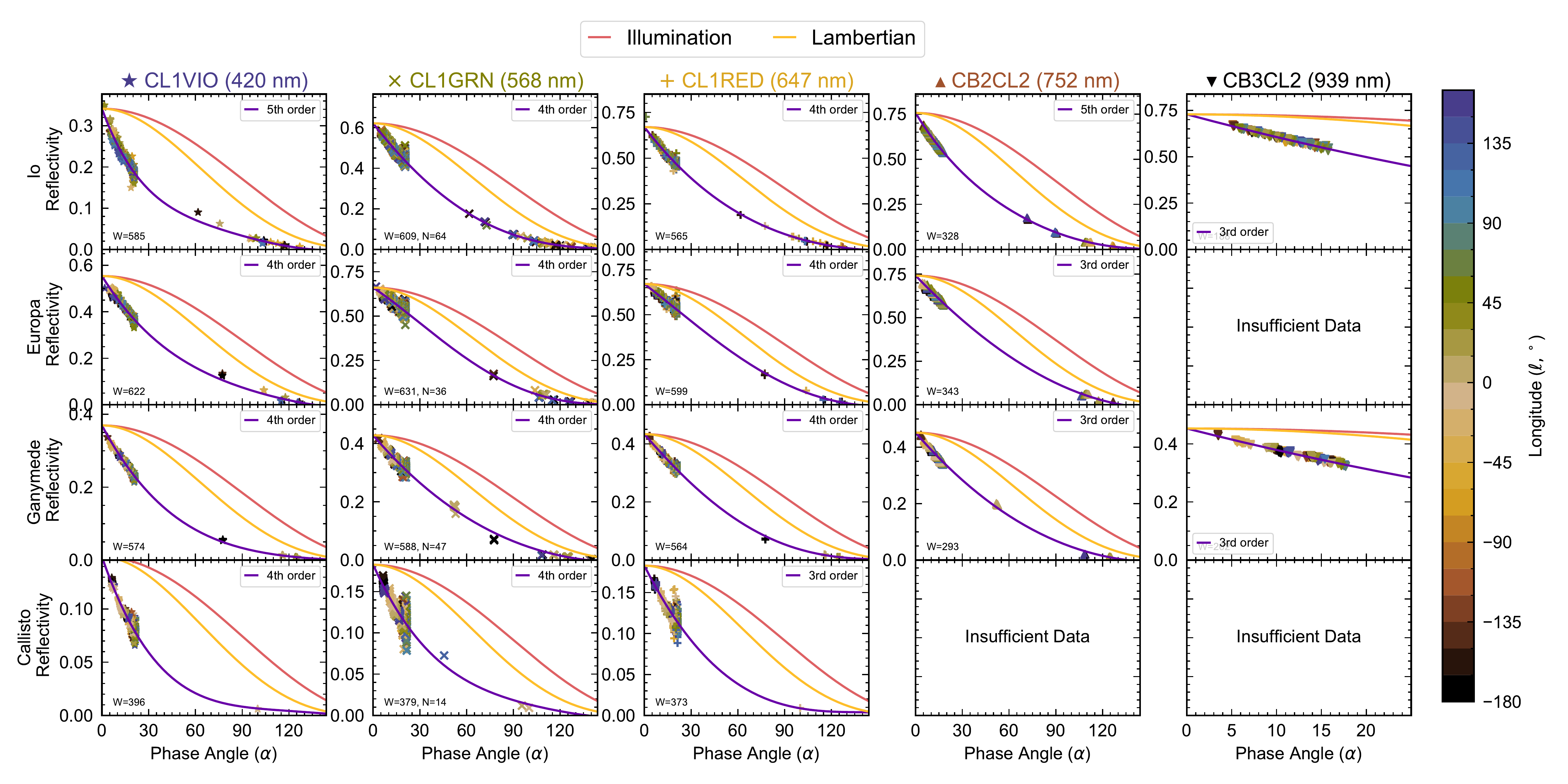}
\caption{Reflectivity as a function of phase angle. Each panel corresponds to a given moon-filter combination. We color each data point by planetocentric longitude and show the best fits as established by the BIC. Purple lines correspond to our best fit polynomial. Orange lines are a Lambertian phase curve normalized to the best fit value at $\alpha$=0 and pink lines are the percentage of the moon that was is illuminated at each phase angle for reference. The phase curves are clearly non-Lambertian. \label{fig:phasecurves}}
\end{figure*}

We show the phase curves for all moons in \autoref{fig:phasecurves} with the best fit polynomial and compare it against the illumination fraction and a Lambertian phase curve, both of which have been normalized to the value of the best fit polynomial at $\alpha$=0\degr{}, i.e. the geometric albedo. A Lambertian phase curve is defined as
\begin{equation}
\mathcal{L}(\alpha) = \frac{\sin(\alpha)+(\pi-\alpha)\cos(\alpha)}{\pi}
\end{equation}
\autoref{tbl:polyfits} tabulates the best fit polynomials. We note that using the tabulated results differs from the machine precision solution by less than 0.05\%. We find that all of the observed phase curves are significantly darker at all phase angles than that of a Lambertian phase curve.

\begin{deluxetable*}{llL}
\tabletypesize{\scriptsize}
\tablecaption{The tabulated best fit polynomial to the phase curves.\tablenotemark{a}\label{tbl:polyfits}}
\tablehead{\colhead{Moon} & \colhead{Filter} & \colhead{$f(\alpha)$}}
\startdata
\multirow{3}{*}{Io} & CL1VIO & 0.341370-1.032816\times10$^{-2}$\alpha+1.658435\times10$^{-4}$\alpha^2-1.522074\times10$^{-6}$\alpha^3+7.031275\times10$^{-9}$\alpha^4-1.249025\times10$^{-11}$\alpha^5\\
 & CL1GRN & 0.621452-9.898741\times10$^{-3}$\alpha+4.331736\times10$^{-5}$\alpha^2+1.745495\times10$^{-8}$\alpha^3-3.288270\times10$^{-10}$\alpha^4\\
 & CL1RED & 0.670283-1.115061\times10$^{-2}$\alpha+6.532755\times10$^{-5}$\alpha^2-1.729996\times10$^{-7}$\alpha^3+2.188850\times10$^{-10}$\alpha^4\\
 & CB2CL2 & 0.756797-1.420334\times10$^{-2}$\alpha+1.295066\times10$^{-4}$\alpha^2-9.210421\times10$^{-7}$\alpha^3+4.440832\times10$^{-9}$\alpha^4-8.926980\times10$^{-12}$\alpha^5\\
 & CB3CL2 & 0.728667-1.301458\times10$^{-2}$\alpha+7.742211\times10$^{-5}$\alpha^2-1.533886\times10$^{-7}$\alpha^3\\
\hline
\multirow{3}{*}{Europa} & CL1VIO & 0.553986-1.080960\times10$^{-2}$\alpha+9.656056\times10$^{-5}$\alpha^2-4.947030\times10$^{-7}$\alpha^3+1.095093\times10$^{-9}$\alpha^4\\
 & CL1GRN & 0.660793-5.872271\times10$^{-3}$\alpha-4.206296\times10$^{-5}$\alpha^2+5.690905\times10$^{-7}$\alpha^3-1.486222\times10$^{-9}$\alpha^4\\
 & CL1RED & 0.671382-6.617609\times10$^{-3}$\alpha-1.859836\times10$^{-5}$\alpha^2+3.371624\times10$^{-7}$\alpha^3-8.032902\times10$^{-10}$\alpha^4\\
 & CB2CL2 & 0.741602-9.289342\times10$^{-3}$\alpha+2.538278\times10$^{-5}$\alpha^2+1.880550\times10$^{-8}$\alpha^3\\
\hline
\multirow{3}{*}{Ganymede} & CL1VIO & 0.369516-7.996506\times10$^{-3}$\alpha+6.939563\times10$^{-5}$\alpha^2-2.883812\times10$^{-7}$\alpha^3+4.795368\times10$^{-10}$\alpha^4\\
 & CL1GRN & 0.428328-6.268761\times10$^{-3}$\alpha+2.914738\times10$^{-5}$\alpha^2-5.598929\times10$^{-8}$\alpha^3+7.891153\times10$^{-11}$\alpha^4\\
 & CL1RED & 0.433450-6.217511\times10$^{-3}$\alpha+1.625945\times10$^{-5}$\alpha^2+1.056810\times10$^{-7}$\alpha^3-4.359767\times10$^{-10}$\alpha^4\\
 & CB2CL2 & 0.451342-6.327110\times10$^{-3}$\alpha+2.484652\times10$^{-5}$\alpha^2-2.004107\times10$^{-8}$\alpha^3\\
 & CB3CL2 & 0.453006-7.840016\times10$^{-3}$\alpha+4.510538\times10$^{-5}$\alpha^2-8.628336\times10$^{-8}$\alpha^3\\
\hline
\multirow{3}{*}{Callisto} & CL1VIO & 0.149008-4.221449\times10$^{-3}$\alpha+4.780409\times10$^{-5}$\alpha^2-2.453344\times10$^{-7}$\alpha^3+4.695099\times10$^{-10}$\alpha^4\\
 & CL1GRN & 0.182535-4.198701\times10$^{-3}$\alpha+4.260496\times10$^{-5}$\alpha^2-2.216542\times10$^{-7}$\alpha^3+4.628491\times10$^{-10}$\alpha^4\\
 & CL1RED & 0.182846-3.784310\times10$^{-3}$\alpha+2.674916\times10$^{-5}$\alpha^2-6.320749\times10$^{-8}$\alpha^3\\
\enddata
\tablenotetext{{\rm a}}{As tabulated, results differ by 0.05\% or less when used at phase angles from 0\degr{}--130\degr{}. We do not recommend the use these fits beyond the range constrained by the data, particularly at high phase angles beyond 140 where no data exists in any filter.}
\end{deluxetable*}

Since we have no constraining data beyond roughly 140\degr{} we do not recommend using these fits at higher phase angles. In general, we find the best fits are polynomials of 3rd to 5th degree. Additionally it can be seen in \autoref{fig:phasecurves} that the VIO filter is particularly difficult to fit across the range of low phase angles ($\alpha < 90\degr{}$). This is due to the steep shape of the fall off at very low phase angles ($\alpha < 20\degr{}$) such that fitting the constraining values from 60\degr{}$<\alpha<$ 90\degr{} becomes difficult. The shutter speed inconsistency is also very notable in GRN and RED filters where there is an increase in scatter near 20\degr{}.

\subsection{High Phase Angles}
\label{ssec:highphase}
\begin{figure*}
\plotone{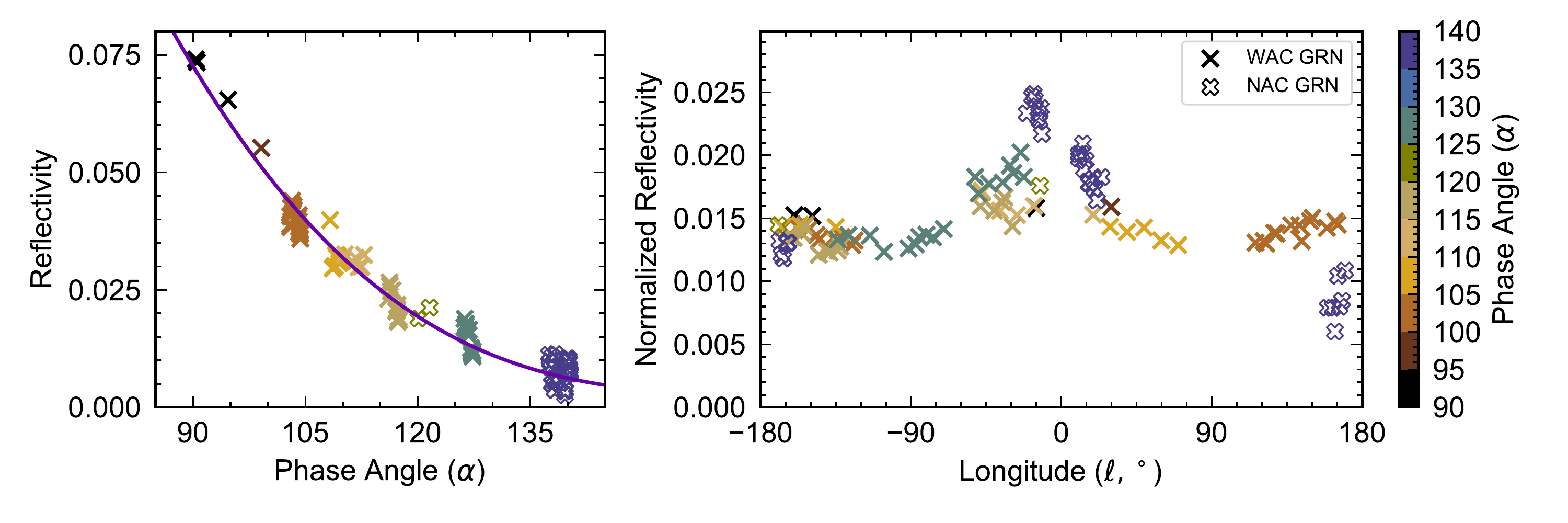}
\caption{Reflectivity of Io in the GRN filter at high phase angles. Data from the WAC is shown in filled crosses and data from the NAC in open crosses. Left: the reflectivity as a function of phase angle and the best fit orbital phase curve in purple. Right: the reflectivity as a function of planetocentric longitude corrected to $\alpha$ = 125\degr{}. Io appears brightest at these phase angles when viewed from  $\ell=0\degr{}$ because the illuminated surface is actually from longitudes roughly 50\degr{} away. \label{fig:highphase}}
\end{figure*}

\begin{figure*}
\plotone{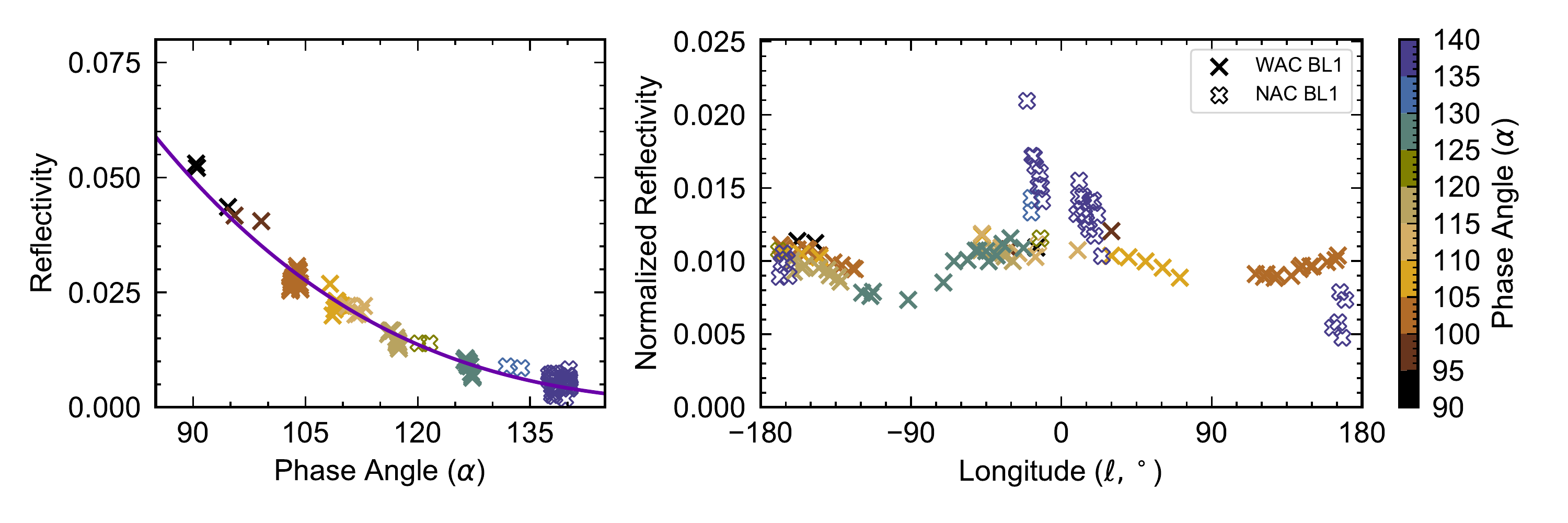}
\caption{Reflectivity of Io in the BL1 filter at high phase angles. Data from the WAC is shown in filled crosses and data from the NAC in open crosses. Left: the reflectivity as a function of phase angle and the best fit orbital phase curve in purple. Right: the reflectivity as a function of planetocentric longitude corrected to $\alpha$ = 125\degr{}. Io appears brightest at these phase angles when viewed from  $\ell=0\degr{}$ because the illuminated surface is actually from longitudes roughly 50\degr{} away \label{fig:blue}}
\end{figure*}

High phase angle observations of exoplanets provide an additional opportunity to map the planet at higher spatial resolution as the planet rotates. There are insufficient data to make rotational light curves at high phase angles for all moons in all filters. We re-examine all six filters and evaluate the planetocentric longitude coverage when including data from phase angles, $\alpha>$90\degr{} and find that when we correct for phase angle variations using our best fit polynomials, Io may have sufficient data in BL1 and GRN with over 100 images between the two cameras for these high phase angles, but the BL1 filter has insufficient data at low phase angles to make a comparison. The reflectivity of Io in the GRN filter at these high phase angles is shown in \autoref{fig:highphase} and in the BL1 filter in \autoref{fig:blue}. The data exhibit a nearly contiguous stretch of longitudinal coverage near $\alpha$ = 125\degr{} and thus we choose to correct the data to this phase angle rather than the smallest phase angle in the range as done previously.

There appears to be a misalignment between the NAC and the WAC data in \autoref{fig:highphase} and \autoref{fig:blue}. The highest phase angle data is provided by the NAC in both. We expect that the rotational variations of a crescent object are more pronounced and dependent on the reflectivity of surface features. When these variations are scaled up to a brighter phase angle they are likely over-exaggerated in comparison to the lower phases. Note that the NAC also provides a few data points near $\alpha$=120\degr{} that exhibits no misalignment with the WAC data. For these reasons the misalignment of the data where $\alpha\sim$140\degr{} is unsurprising and suggests a physical explanation rather than a camera calibration issue. This suggests that the illuminated surfaces are strongly directionally scattering.

\begin{figure*}
\raggedright
\hspace{0.2cm}
\includegraphics[width=0.32\textwidth]{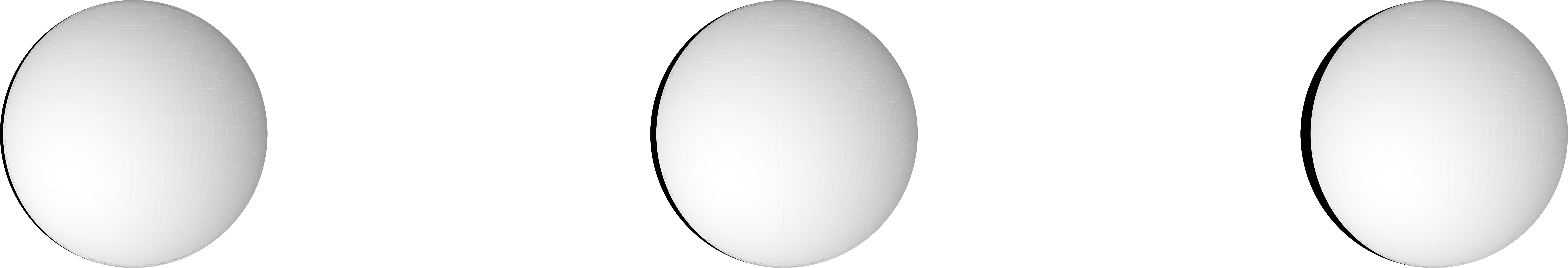} \hspace{3cm}
\includegraphics[width=0.32\textwidth]{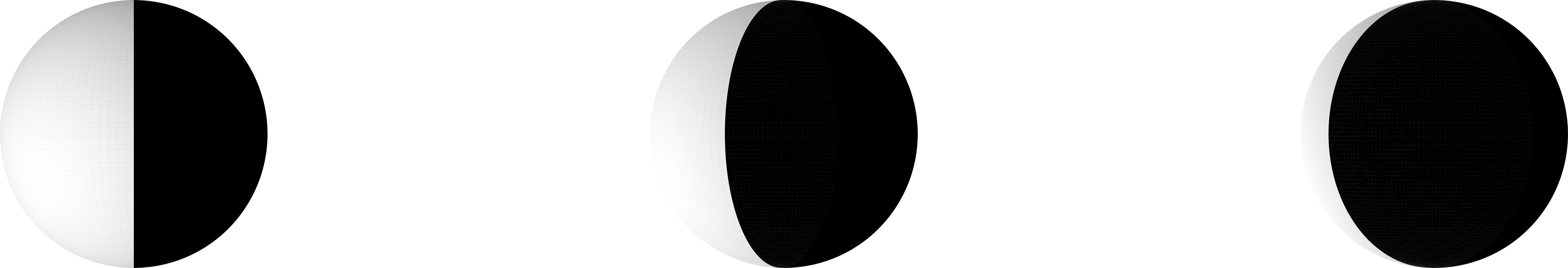}\\
\centering
\includegraphics[width=0.32\textwidth]{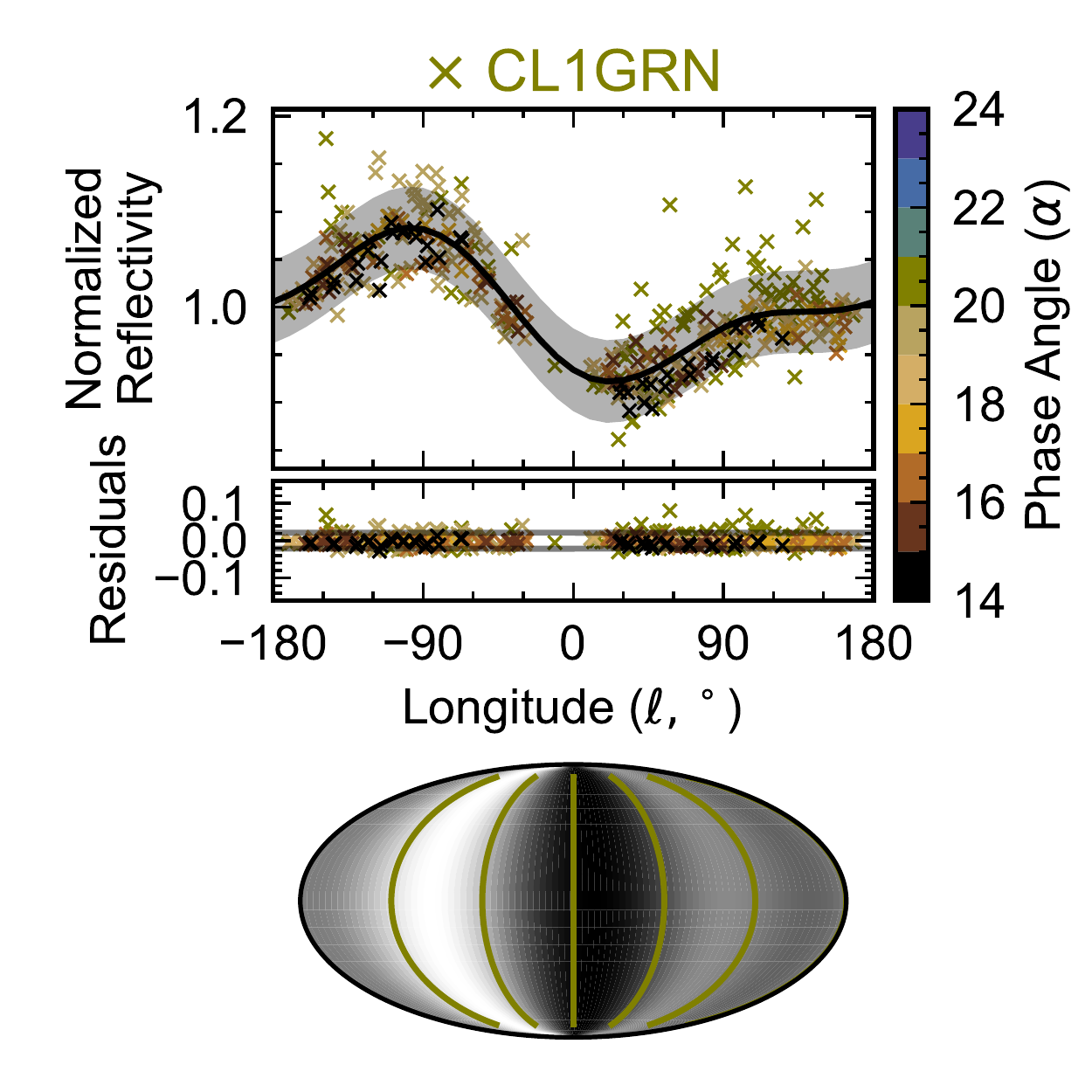}
\includegraphics[width=0.32\textwidth]{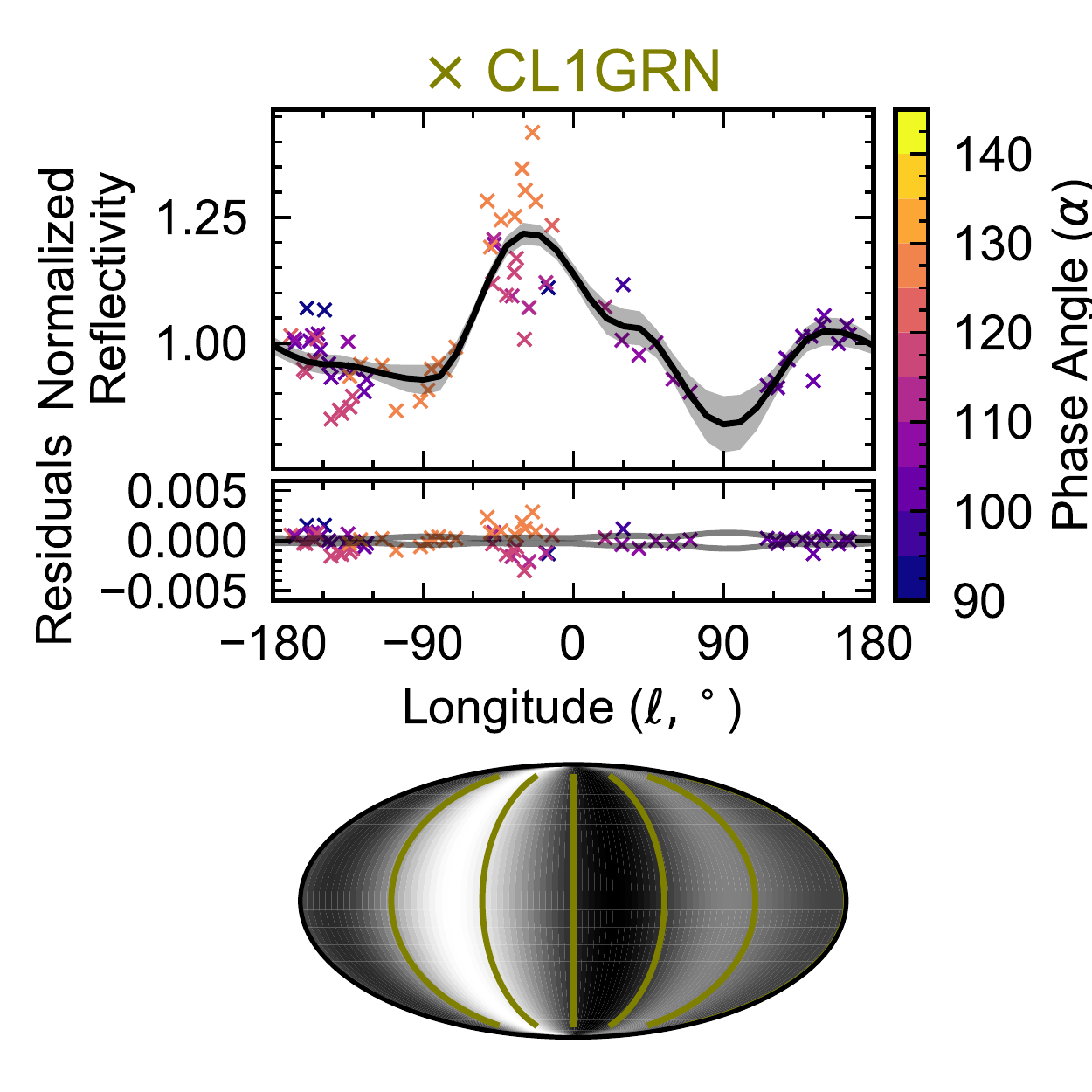}
\includegraphics[width=0.32\textwidth]{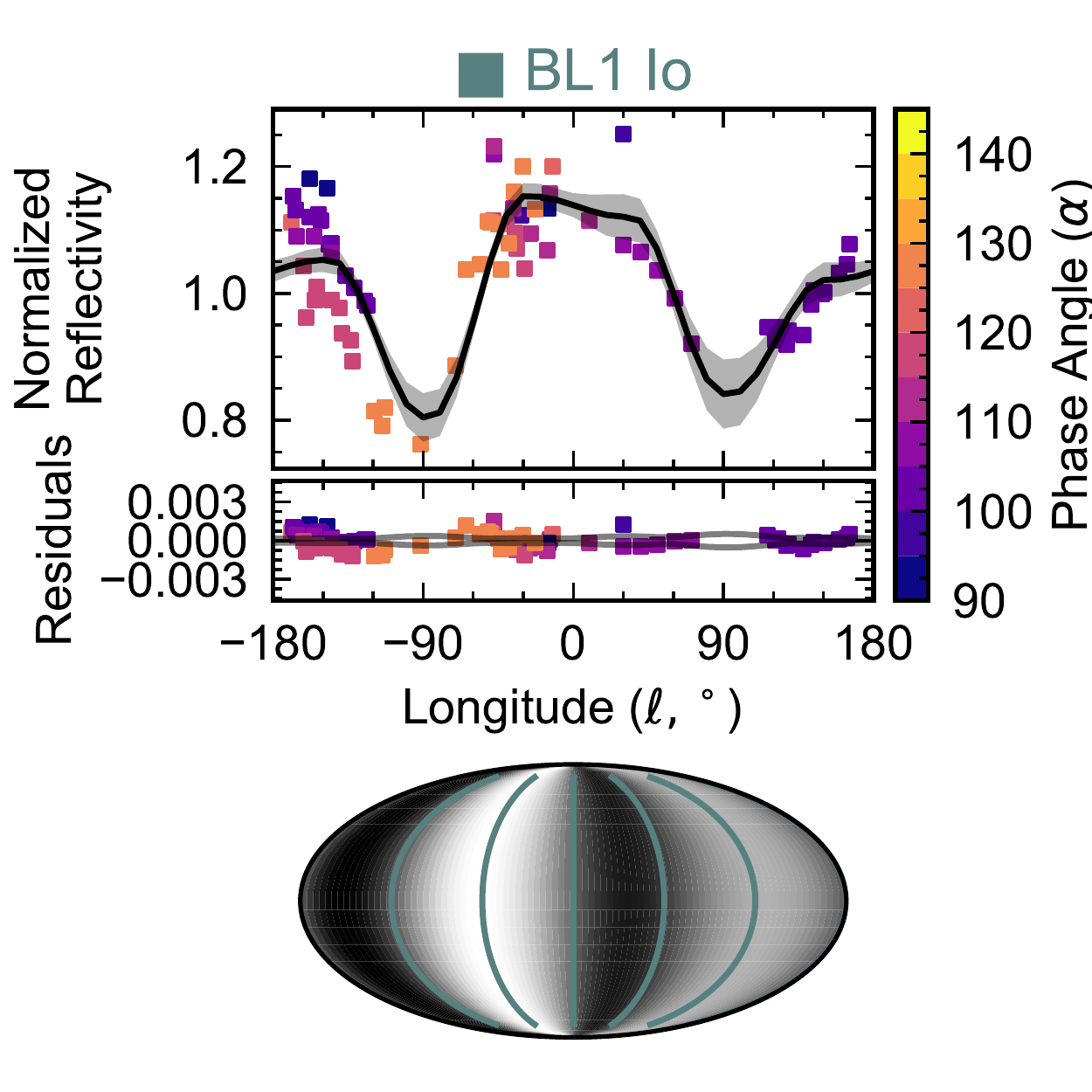}\\
\caption{A comparison of the low phase angle and high phase angle rotational light curve of Io. Left: Using phase angles between \parange{} in the GRN filter. Middle: Using phase angles between 90\degr{}--130\degr{} in the GRN filter. Right: Using phase angles between 90\degr{}--130\degr{} in the BL1 filter. The cartoons across the top shows the smallest, mean, and largest phase angle included in the range. Note again the shift in the brightness corresponding to the higher phase angle and longitude of view and yet the maps are consistent. The panels are the same as in \autoref{fig:io} \added{except we show the reflectivity normalized by the average}. \label{fig:phasecomp}}
\end{figure*}

We show a comparison of the data, fit, residuals, and \emph{inverted map} (using the same N=6 number of slices) from both low phase angles (the middle panels from \autoref{fig:io}) and high phase angles in \autoref{fig:phasecomp} using only WAC data. Because the variation at these high phase angles is so extreme in comparison to the shorter phase angle data provided by the WAC, we omit NAC data in this range to model the variations. We include a cartoon diagram of the moon as illuminated in the phase angle range considered for both rotational light curve inversions. Note that when the planetocentric longitude is 0\degr{} \replaced{when}{and} the phase angle is 140\degr{}\added{,} the planetocentric longitudes of the surface features that are illuminated are 50\degr{}--90\degr{} away \added{from the sub-spacecraft longitude. In contrast,} \deleted{vs }when the phase angle is 90\degr{}\added{,}\deleted{ where} the illuminated surface features are \deleted{from} 0\degr{}--90\degr{} away.

We quantify the variation as in \autoref{ssec:rotational} and find that at high phase angles the variation is larger (in GRN, 38 $\pm$ 6\%), nearly double the percent variation at low phase angles. The shape of the rotational variations is also different and results in a similar map. However, since we assume surfaces are Lambertian scatterers and the phase angle range is quite wide, the \emph{inverted map} is not necessarily robust since we do not know the true scattering properties of the surface and it is difficult to correlate with surface features. For example, if one observes the moon from above 0$\degr{}$ in latitude and planetocentric longitude, at a phase angle of roughly 179\degr{}, the reflectivity variations are due to the sliver of surfaces at a planetocentric longitude of roughly $\pm$90\degr{} \added{and if these surfaces have a strongly directional scattering component our model cannot capture this}.

\subsection{Color}
\label{ssec:color}
With direct imaging missions we will have access to several filters or spectral windows. It has been proposed that color can be used to determine surface properties of terrestrial-type planets \citep{Hegde2013, Fujii2017}, detect exomoons \citep{Agol2015}, and mitigate the contamination from background objects \citetext{Exo-S Final Report}. It is also important to consider surface color variations as contaminants in classification schemes and that we \added{do} not confuse these variations as being from an atmospheric source.

\begin{figure*}
\plotone{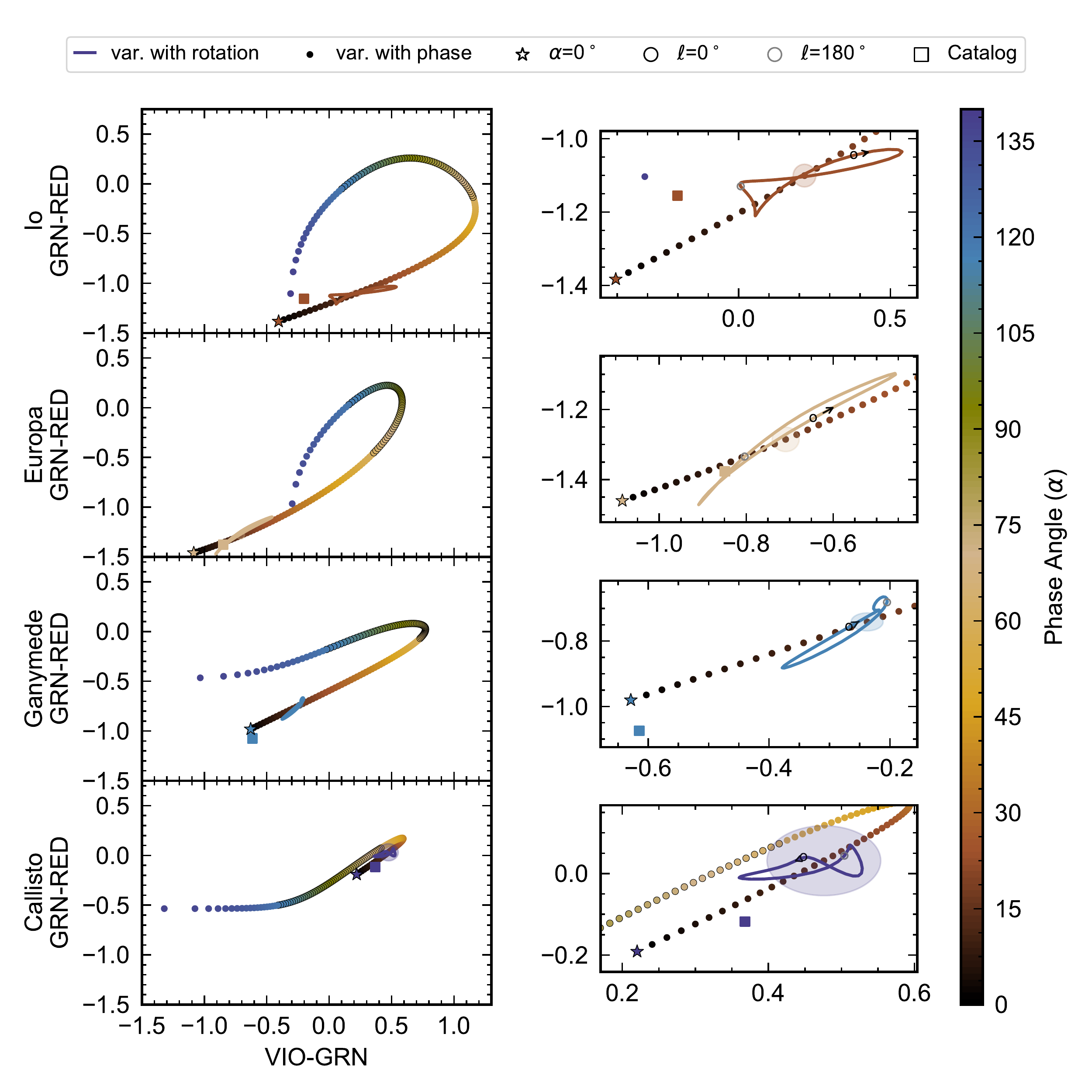}
\caption{Variation in color-color of the Galilean satellites in (VIO--GRN), (GRN--RED). From top to bottom: Io, Europa, Ganymede, and Callisto. The left panels show the variations on the same scale. The right panels are zoom-ins on the low phase angle region. The color variations with phase angle as fit are shown as colored circles every degree of phase \added{with the direct imaging phase angles of interest denoted with black outlines}. The stars mark $\alpha$=0\degr{}. The color variations with planetocentric longitude are shown as fit from $\alpha$=\parange and are the solid lines where the black circles mark $\ell$=0\degr{}, the sub-Jupiter planetocentric longitude, with the arrow showing the direction of increasing planetocentric longitude, the gray circles mark the anti-Jupiter planetocentric longitude. We show the typical errors on the rotational variation as an ellipse where the major and minor axis are the 1~sigma errors. We overplot on both panels the filter-integrated colors of the Galilean satellites in squares as provided by \citet{Madden2018}. \label{fig:color}}
\end{figure*}

\begin{figure*}
\plotone{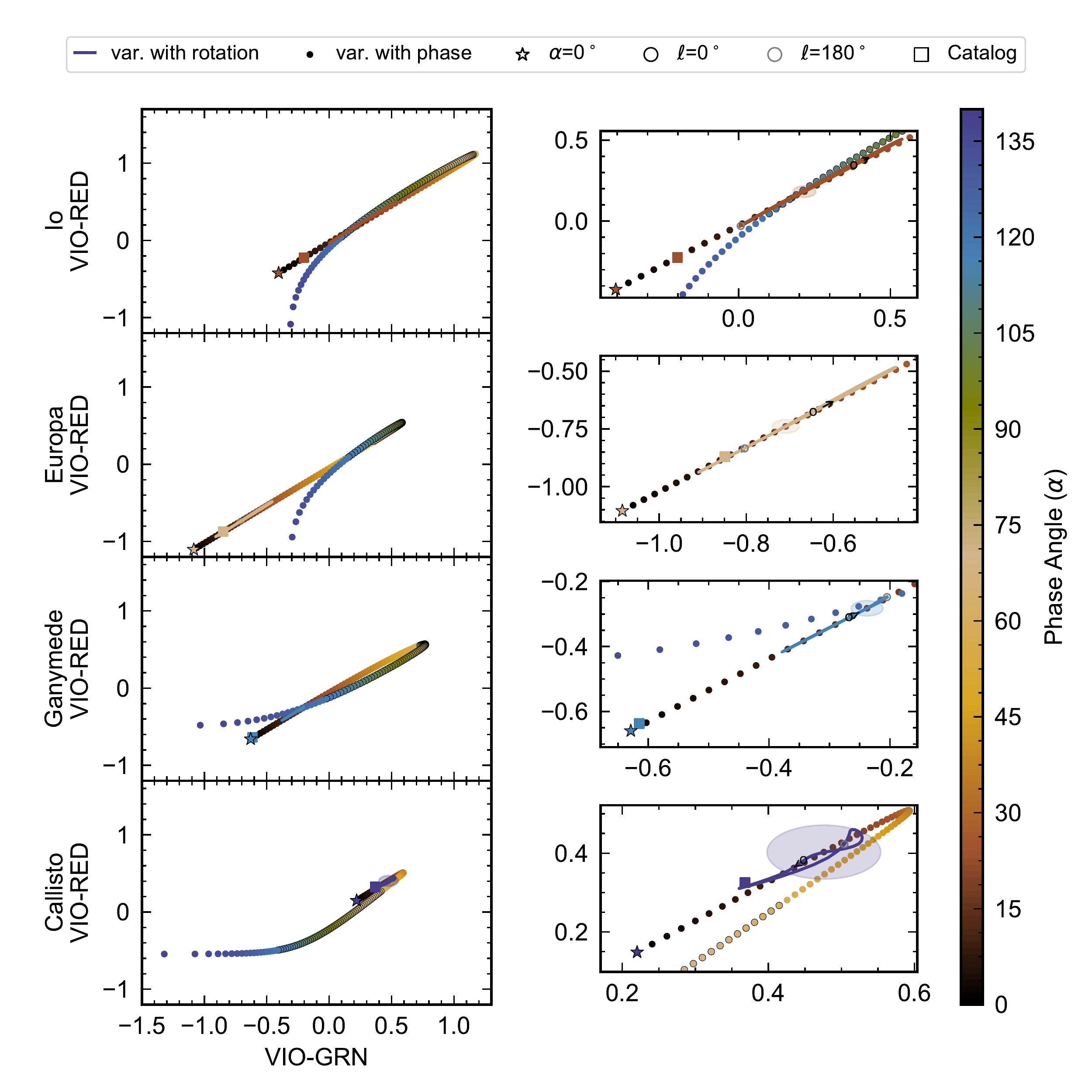}
\caption{Same as \autoref{fig:color} but in (VIO--GRN), (VIO--RED). \label{fig:colorvgvr}}
\end{figure*}

To compute the rotational color-color variations of each moon as a function of planetocentric longitude for the low phase angle range of \parange{}, we follow the prescriptions of \citet{Cahoy2010} and subsequently \citet{Mayorga2016},
\begin{equation}
(m_{f_1} - m_{f_2})(\ell) = -2.5 \log_{10} \left(\frac{A_{f_1}(\ell)}{A_{f_2}(\ell)}\frac{\int_\lambda F_{\odot}(\lambda)T_{f_1}(\lambda) {\rm d}\lambda}{\int_\lambda F_{\odot}(\lambda)T_{f_2}(\lambda) {\rm d}\lambda} \right)
\end{equation}
where $m_{f_1}$ and $m_{f_2}$ are the magnitude in the two filters, $A_{f_1}(\ell)$ is the reflectivity variation in the fit of filter $f_1$ data, $T_{f_1}(\lambda)$ is the flux normalized transmission function, and $F_{\odot}(\lambda)$ is the incident solar flux. To compute the illumination color-color variations we replaced $A_{f_1}(\ell)$ , with $A_{f_1}(\alpha)$. \citet{Cahoy2010} found that ($B-V$), ($R-I$) or ($B-V$), ($V-I$) were best at differentiating extrasolar giant planets. \citet{Hegde2013} additionally used ($B-V$), ($B-I$) for exo-Earth surfaces. Using VIO, GRN, and RED, we show the variation in color with rotation and illumination for the combinations (VIO--GRN), (GRN--RED) and (VIO--GRN), (VIO--RED) color-color diagrams in \autoref{fig:color} and \autoref{fig:colorvgvr}. We also filter-integrate the albedo spectra and display them in \autoref{fig:color} for comparison. The colors we find in comparison to the literature full phase colors are typically in agreement within our error as differences in phase angle, observed hemisphere (and subsequent volcanic activity in the case of Io) can explain the discrepancies. 

The variations in color with illumination are typically more pronounced in (VIO--GRN) or (VIO--RED) with (GRN-RED) showing the smallest variations. These quantities are tabulated in \autoref{tbl:colors}. The color-color variations can be quite large over the entire phase angle range. We also computed the variation over just the phase angle range which direct imaging missions are likely to observe, roughly 60\degr{} to \replaced{130}{120}\degr{}, and find that for some moons in certain filters there is little variation over this range (Europa) and for others the bulk of the variation is exhibited in this range of phase angles (Io). This suggests that measuring and understanding the phase curves of terrestrial exoplanets can be informative for determining surface composition and potentially atmospheric properties.

\begin{deluxetable}{lcCCC}
\tabletypesize{\scriptsize}
\tablecaption{The maximum color variation in magnitudes exhibited by each moon as a function of illumination and rotation.\label{tbl:colors}}
\tablehead{\colhead{Moon} & \colhead{Color} & \colhead{$\alpha$=0\degr{}--140\degr{}} & \colhead{$\alpha$=60\degr{}--120\degr{}} & \colhead{$\ell$}}
\startdata
\multirow{3}{*}{Io} & VIO--GRN & 1.58 & 1.08 & 0.53 \pm 0.11 \\
 & VIO--RED & 3.41 & 1.10 & 0.54 \pm 0.11 \\
 & GRN--RED & 2.55 & 0.41 & 0.2 \pm 0.1 \\
\hline\hline
\multirow{3}{*}{Europa} & VIO--GRN & 1.67 & 0.45 & 0.5 \pm 0.1 \\
 & VIO--RED & 2.83 & 0.48 & 0.5 \pm 0.1 \\
 & GRN--RED & 2.52 & 0.70 & 0.37 \pm 0.10 \\
\hline\hline
\multirow{3}{*}{Ganymede} & VIO--GRN & 3.06 & 0.82 & 0.2 \pm 0.1 \\
 & VIO--RED & 1.23 & 0.73 & 0.2 \pm 0.1 \\
 & GRN--RED & 1.06 & 0.28 & 0.2 \pm 0.1 \\
\hline\hline
\multirow{3}{*}{Callisto} & VIO--GRN & 2.88 & 0.86 & 0.17 \pm 0.21 \\
 & VIO--RED & 1.05 & 0.78 & 0.15 \pm 0.19 \\
 & GRN--RED & 0.71 & 0.59 & 0.08 \pm 0.24 \\
\enddata
\end{deluxetable}

While we are limited by the errors on the fits as to the precise rotational variations in color space, e.g. Callisto, it is notable that we measure very similar colors for Europa, Ganymede, and Callisto in these color spaces. Due to Io's volcanic activity the surface of the moon has been studied at some length \citep[e.g.][etc.]{Simonelli1984, McEwen1985, Simonelli1986a, Simonelli1986b, Simonelli1986c, Simonelli1988, Simonelli2001, Simonelli1994, Domingue1997, Domingue1998, Geissler1999, Williams2011}. Io boasts a wide range of colors that have been postulated to be a function of varying quantities of sulfur compounds and silicates across the surface \citep{Geissler1999}. Volcanically active areas (typically red on Io's surface), such as Pele\footnote{We refer to the maps and names found at \emph{The Gazetteer of Planetary Nomenclature}, https://planetarynames.wr.usgs.gov/} near \(\ell\)=115\degr{}, appear to cause the brightening in RED detected at those longitudes.  The predominantly white regions on Io (such as Bosphorus, Media, Tarsus, and Colchis Regio) correlate with the planetocentric longitudes that lead to higher VIO albedo. Small green-ish yellow deposits, such as Lei-Kung Fluctus and Isum Patera located in the northern hemisphere near 150\degr{} planetocentric longitude, can be correlated with the flatness of the phase curve in the GRN filter at this planetocentric longitude rather than the fall off that the RED filter phase curve exhibits.

\subsection{Comparison with Prior Observations}
\begin{figure}
    \plotone{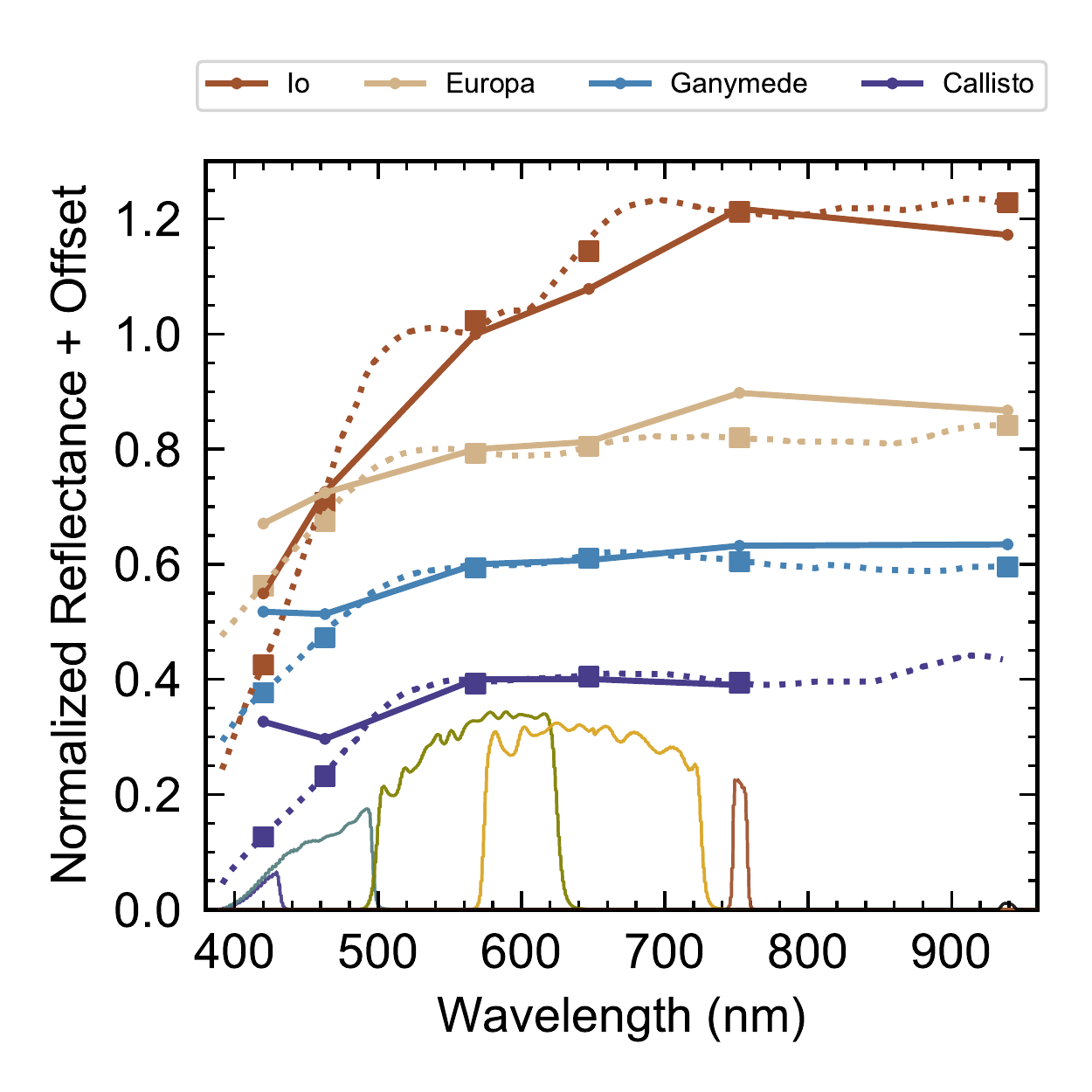}
    \caption{A comparison of the geometric albedos as measured by \cass{}/VIMS and \CISS{}. The solid lines are the geometric albedos as given by our phase curve fits where the points mark the effective wavelength of each filter (the transmission functions are shown across the bottom). The dashed lines are the spectra from \citet{McCord2004} where the squares mark the filter-integrated albedos at the effective wavelength of each filter. \label{fig:geom-spectra}}
\end{figure}

\citet{McCord2004} published the spectra of the Galilean satellites using \cass{}/VIMS-V results and compared against the spectroscopic observations of \citet{Karkoschka1994}. We filter integrate the results of \citet{McCord2004} and compare those against our results in \autoref{fig:geom-spectra}. The \citet{McCord2004} data were published in a figure as normalized reflectance with an offset, where each moon was normalized to a value at 563~nm. Io was normalized to unity, Europa to 0.8, Ganymede to 0.6, and Callisto to 0.4. Our closest filter is GRN with an effective wavelength of 568~nm and we normalize our spectra to the same values at this wavelength. There is a discrepancy between the work here and \citet{McCord2004} in the VIO filter; our measurements suggest the moons are brighter. Phase curve observations of Europa, Ganymede, and Callisto are also published in \citet{McCord2004}, but the dataset is small and sparsely populates the phase angle range with no data at phase angles greater than 115\degr{}.

We can compare our rotational light curves to the work of \citet{Millis1975} (the source for \citet{Fujii2014}), where they published the rotational light curves in UBV magnitudes as a function of orbital phase angle\footnote{Note that this is orbital phase angle, not illumination phase angle and to convert to planetocentric longitude one must subtract 180\degr{}.}. The effective wavelength of V is 551~nm and can be compared against the results from the GRN filter. The shape of the light curves we derive for Io, Europa, and Ganymede are very similar to those of \citet{Millis1975}. The \citet{Millis1975} results for Callisto are more sinusoidal in shape than our results. 

\section{Impact for Observing Exoplanets}
\label{sec:impact}
To simulate direct imaging observations of an exoplanet, we can paint the surfaces of the Galilean satellites onto an Earth-sized exoplanet and compute the variation in the expected planet-to-star flux ratio or contrast due to rotation or illumination. We use the standard equation for the reflected light of a planet, but add in the variations on the albedo with planetocentric longitude and the phase function,
\begin{equation}
\frac{F_{\rm P}}{F_*}\left(\alpha,\ell\right) = \left(\frac{R_{\rm P}}{a}\right)^2 A_{\rm g}\Phi(\alpha)\Psi(\ell),
\end{equation}
where $A_{\rm g}$ is the geometric albedo, the reflectivity at full phase (compared to a Lambertian disk), $R_{\rm P}$ is the radius of the planet, $a$ is the orbital semi-major axis of the planet, $\Phi(\alpha)$ is the phase function or the phase curve normalized to unity, and $\Psi(\ell)$ is the longitudinal or rotational phase function. As we saw in \autoref{ssec:highphase}, Io has a very different rotational variation at high phase angles than at low phase angles. In reality, there would be another term to account for the scattering properties of the different surfaces on the moon. As written, the equation to reproduce the phase curve data is $A_{\rm g}\Phi(\alpha)\Psi(\ell, \alpha)$. We refer to the quantity $\left(\frac{F_{\rm P}}{F_*}\right)$ as $C$ or the contrast and we will give values in parts per billion (ppb).

\begin{figure*}
    \plotone{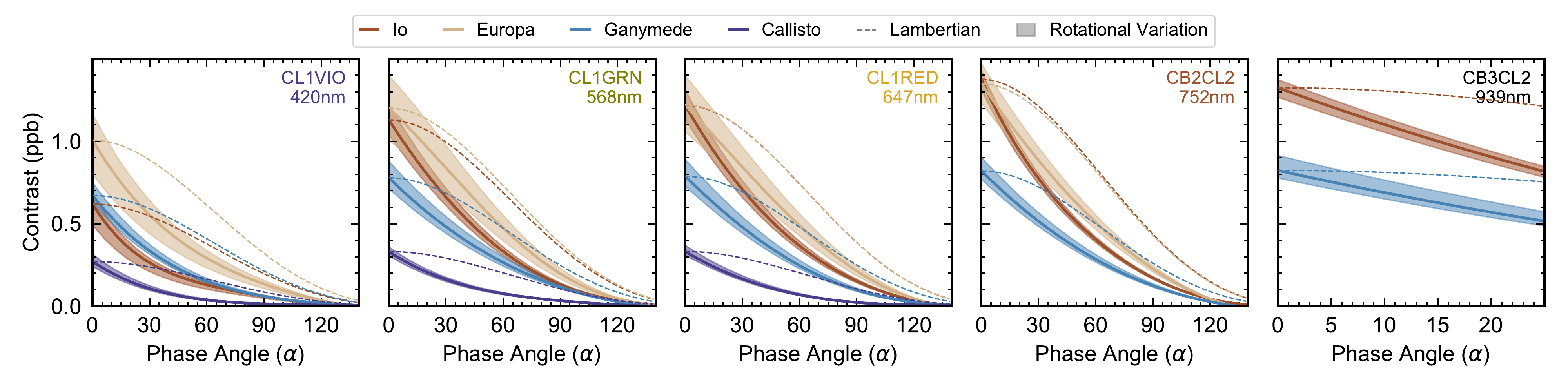}
    \caption{The computed contrast curves of each Galilean satellite as if it were an Earth-sized exoplanet at 1~AU from a Sun-like star. The solid lines correspond to the contrast as computed from our phase curve fit and the shaded regions correspond to the variation potentially caused by rotational reflectivity changes (assuming the low phase angle fit is true everywhere). The dashed lines show the contrast as expected from a Lambertian phase curve with the same geometric albedo. \label{fig:contrast}}
\end{figure*}

We take our phase curve fits in \autoref{fig:phasecurves} and the low phase angle rotational fits in \autoref{fig:io}-\autoref{fig:callisto} as the potential range in reflected light brightness at a fixed phase angle resulting from surface variations (higher phase angles are likely to vary more than this, see again \autoref{ssec:highphase}) and compute the expected contrast for each of the Earth-sized Galilean satellites at 1~AU. We show these results for the VIO, GRN, RED, CB2, and CB3 filters in \autoref{fig:contrast} as compared to a Lambertian phase curve with the same geometric albedo.

By design, direct imaging observations occult the on-sky region around the star. Thus, observations are constrained to the phase angles near quadrature, 60\degr{}$<\alpha<$120\degr{}. At 60\degr{} the Galilean satellites are already 40\% as bright as at full phase or less.

For rotational variations, at low phase angles in the GRN filter, the rotational variations of Io are of order 0.14~ppb (a 16\% flux variation) peak to peak. At high phase angles, the Io-like Earth-sized planet at 1~AU would vary 0.009~ppb (a 38\% flux variation) in GRN and 0.006~ppb in BL1 (a 35\% flux variation). For reference, \citep{TheLUVOIRTeam2019}, which hopes to detect the Earth at 10~parsecs, will require on the order of 0.1~ppb precision. We have shown that these rotational variations will require precision on the order of 0.1~ppb to create brightness maps if low phase angles are viewed. However, the direct imaging bias toward higher phase angles will mean their variations will be of order roughly 0.001~ppb.

If we assume the low phase angle rotational variation is the case at all phase angles, in GRN an observation of an exo-Earth with a Callisto-like surface will need to detect a 0.077~ppb signal at 60\degr{} and 0.034~ppb at 90\degr{}. Without high phase angle rotational variation we can only assume that longitudinal variations will be of order 16\%, i.e. 0.01~ppb and 0.001~ppb respectively. Europa has the largest longitudinal variations and is much brighter than the other satellites, which leads to a 0.474~ppb signal at 60\degr{} and 0.198~ppb at 90\degr{}. Europa's rotational variations will only require 0.1~ppb and 0.01~ppb respectively.

The contrast, ratios, and variation with illumination and rotation in each color and/or filter is summarized for each moon in \autoref{tbl:results}. Note BL1 has not been fit iteratively and results are reported with less precision since they are based off of the initial 3rd order polynomial.

\begin{deluxetable*}{lcRRRRR}
\tabletypesize{\small}
\tablecaption{The expected contrast for each moon by color and filter assuming it is an Earth-sized planet at 1~AU from a Sun-like star and the percentage variation in rotation expected from low phase angle data. \label{tbl:results}}
\tablehead{\colhead{Moon} & \colhead{Filter} & \colhead{C(0\degr{})} & \colhead{C(60\degr{})} & \colhead{C(90\degr{})} & \colhead{C(120\degr{})} & \colhead{Var. with $\ell$} \\
\cline{3-7} & & ({\rm ppb}) & ({\rm ppb}) & ({\rm ppb}) & ({\rm ppb}) & (\%) \\ 
\cline{7-7} & & & & & & {\rm 14\degr{}-24\degr{}}}
\startdata
\multirow{6}{*}{Io} & CL1VIO & 0.62 & 0.13(21\%) & 0.06(10\%) & 0.01(2\%) & 31 \pm 8 \\ 
 & BL1\tablenotemark{a} & 0.8 & 0.2(30\%) & 0.1(10\%) & 0.0(0\%) & \nodata \\ 
 & CL1GRN & 1.13 & 0.33(29\%) & 0.13(12\%) & 0.04(3\%) & 16 \pm 6 \\ 
 & CL1RED & 1.22 & 0.37(30\%) & 0.15(13\%) & 0.04(3\%) & 14 \pm 5 \\ 
 & CB2CL2 & 1.38 & 0.40(29\%) & 0.17(13\%) & 0.04(3\%) & 8 \pm 1 \\ 
 & CB3CL2 & 1.32 & 0.35(27\%) & 0.13(10\%) & 0.03(2\%) & 8 \pm 3 \\ 
\hline\hline
\multirow{6}{*}{Europa} & CL1VIO & 1.01 & 0.29(29\%) & 0.14(13\%) & 0.04(4\%) & 36 \pm 5 \\ 
 & BL1\tablenotemark{a} & 1.1 & 0.4(30\%) & 0.2(20\%) & 0.0(0\%) & \nodata \\ 
 & CL1GRN & 1.20 & 0.47(39\%) & 0.20(16\%) & 0.05(4\%) & 33 \pm 7 \\ 
 & CL1RED & 1.22 & 0.49(40\%) & 0.21(18\%) & 0.05(4\%) & 28 \pm 6 \\ 
 & CB2CL2 & 1.35 & 0.51(38\%) & 0.23(17\%) & 0.05(3\%) & 18 \pm 2 \\ 
 & CB3CL2 & 1.3 & \nodata & \nodata & \nodata & \nodata \\ 
\hline\hline
\multirow{6}{*}{Ganymede} & CL1VIO & 0.67 & 0.15(23\%) & 0.06(9\%) & 0.02(3\%) & 18 \pm 4 \\ 
 & BL1\tablenotemark{a} & 0.7 & 0.2(30\%) & 0.1(10\%) & 0.0(0\%) & \nodata \\ 
 & CL1GRN & 0.78 & 0.27(34\%) & 0.12(15\%) & 0.03(4\%) & 21 \pm 5 \\ 
 & CL1RED & 0.79 & 0.25(31\%) & 0.10(12\%) & 0.02(3\%) & 21 \pm 5 \\ 
 & CB2CL2 & 0.82 & 0.29(35\%) & 0.12(15\%) & 0.03(3\%) & 16 \pm 2 \\ 
 & CB3CL2 & 0.82 & 0.23(28\%) & 0.09(11\%) & 0.02(3\%) & 17 \pm 2 \\ 
\hline\hline
\multirow{6}{*}{Callisto} & CL1VIO & 0.27 & 0.04(14\%) & 0.01(6\%) & 0.01(3\%) & 27 \pm 11 \\ 
 & BL1\tablenotemark{a} & 0.2 & 0.1(30\%) & \nodata & \nodata & \nodata \\ 
 & CL1GRN & 0.33 & 0.08(23\%) & 0.03(10\%) & 0.01(3\%) & 16 \pm 17 \\ 
 & CL1RED & 0.33 & 0.07(21\%) & 0.02(7\%) & 0.01(3\%) & 18 \pm 14 \\ 
 & CB2CL2 & 0.3 & \nodata & \nodata & \nodata & \nodata \\ 
 & CB3CL2 & \nodata & \nodata & \nodata & \nodata & \nodata \\ 
\enddata
\tablenotetext{\rm a}{Note that BL1 is not fit iteratively with rotational variations removed and thus is an unrefined approximation where data is available from a simple 3rd order polynomial fit. There is insufficient data at low phase angles to allow for this.}
\end{deluxetable*}

\section{Conclusions}
\label{sec:conc}
We have presented here an exoplanet focused analysis of \CISS{} images taken during the flyby of Jupiter in late 2000 where the Galilean satellites appeared. Using navigation data along with computer vision techniques, we accurately located the moons' positions in the images and conducted aperture photometry to determine the reflectivity of each satellite as a function of longitude and illumination phase angle. We modeled the rotation variations using \verb|planetSlicer|, which breaks down the planetary body into a number of slices to determine the brightness at a phase angle and longitude that would produce the variations we see. We modeled the phase curves, the illumination variations, with simple polynomials. Using these fits, we computed the color variation each moon exhibits as a function of longitude and illumination. We publish our fits for the community to use, as well as all of our reduced data.

We find that the amplitude of the rotational variations ranges from 8--36\% depending on the filter with which the moons are observed. For all moons but Callisto, which had the noisiest data and fewest observed rotations, we recover the rotations periods to within 1\%. All of the moons are significantly darker than predicted by a Lambertian phase function with steep decreases at low phase angles. While there is insufficient data for all moons, Io data suggests that at high phase angles \added{fractional} rotational variations may be larger than at low phase angles. Using our fits for the phase curves and rotational variations we measured the color variations as a function of illumination and rotation, finding that the colors of the moons can vary by magnitudes across all phase angles. To detect an Earth-sized Io at 1~AU from a Sun-like star would require precisions of order 0.1~ppb in contrast to detect at 60\degr{} illumination phase angle and rotational variations will require a precision of 0.01~ppb. An object like Callisto would be much fainter. Should instrumentation be precise enough to isolate the reflected light from an exoplanet, our analysis of the varied surfaces of the Galilean satellites indicate that surface compositional variations can be inferred from brightness and color variations over the course of a planet's rotation.

Since the Galilean satellites have little to no atmosphere we have been able to isolate reflectivity variations as a function of illumination and rotation from any atmospheric variations. If they had thick atmospheres and/or bright water clouds like Earth, such variations would be entangled with the temporal and spatial variations due to weather. It is important to consider the potential contamination into the reflected or emitted light signal of a terrestrial exoplanet by the surface.

Future flagship missions such as, \emph{LUVOIR}, \emph{HabEx}, and beyond, will be able to image warm and cool Earth-sized planets. Through long term monitoring, rotational and orbital light curves will need to be interpreted to infer underlying surface and weather phenomena. We have demonstrated here that Earth-sized exoplanets with Galilean-like surfaces will undergo potentially detectable contrast variations with illumination angle and planetocentric longitude such that we can begin to characterize their surfaces and disentangle their atmospheres from a reflected light observation.

\acknowledgements
We thank N.L. Mathes for his insights into computer vision that greatly improved the robustness of the photometric pipeline as well as his support in the home stretches in the final preparation of this manuscript. We also thank M.S. Marley for successfully not being too annoyed by our questions, usually.
Work performed by L.C.M. was supported by the Harvard Future Faculty Leaders Postdoctoral fellowship. D.P.T. acknowledges support from NASA Exoplanets Research Program grant NNX16AB49G.
This publication was made possible through the support of a grant from the John Templeton Foundation. The opinions expressed in this publication are those of the authors and do not necessarily reflect the views of the John Templeton Foundation.

\facilities{Planetary Data System Cartography and Imaging Science Node, Planetary Data System Ring-Moon Systems Node, The Navigation and Ancillary Information Facility,}

\software{Astropy \citep{Astropy}, NumPy \citep{NumPy}, Matplotlib \citep{matplotlib}, PlanetSlicer \citep{PlanetSlicer}, CISSCAL \citep{Knowles2016}, Pandas \citep{Pandas}, SpiceyPy \citep{SpiceyPy}, SciPy \citep{SciPy}, skimage \citep{skimage}, vicar \citep{vicar}, Matplotlib \citep{matplotlib}, Uncertainties \citep{uncertainties}}

\bibliographystyle{aasjournal}
\bibliography{Mendeley}

\begin{thebibliography}{}
\expandafter\ifx\csname natexlab\endcsname\relax\def\natexlab#1{#1}\fi
\providecommand{\url}[1]{\href{#1}{#1}}

\bibitem[{Acton {et~al.}(2018)Acton, Bachman, Semenov, \& Wright}]{Acton2018}
Acton, C., Bachman, N., Semenov, B., \& Wright, E. 2018, Planetary and Space
  Science, 150, 9

\bibitem[{Agol {et~al.}(2015)Agol, Jansen, Lacy, Robinson, \&
  Meadows}]{Agol2015}
Agol, E., Jansen, T., Lacy, B., Robinson, T.~D., \& Meadows, V. 2015, The
  Astrophysical Journal, 812, 5.
\newblock
  \url{http://stacks.iop.org/0004-637X/812/i=1/a=5?key=crossref.0f64fe7acf335175f8904359e04b6167}

\bibitem[{Annex(2017)}]{SpiceyPy}
Annex, A. 2017, 3rd Planetary Data Workshop 2017, 2017, 2.
\newblock
  \url{https://github.com/AndrewAnnex/SpiceyPy%0Ahttps://ui.adsabs.harvard.edu/#abs/2017LPICo1986.7081A}

\bibitem[{Becker {et~al.}(2001)Becker, Archinal, Colvin, Davies, Gitlin, Kirk,
  \& Weller}]{Mosaics}
Becker, T., Archinal, B., Colvin, T., {et~al.} 2001, in Lunar and Planetary
  Inst.{\~{}}Technical Report, Vol.~32, Lunar and Planetary Science Conference

\bibitem[{Brown {et~al.}(2003)Brown, Baines, Bellucci, Bibring, Buratti,
  Capaccioni, Cerroni, Clark, Coradini, Cruikshank, Drossart, Formisano,
  Jaumann, Langevin, Matson, McCord, Mennella, Nelson, Nicholson, Sicardy,
  Sotin, Amici, Chamberlain, Filacchione, Hansen, Hibbitts, \&
  Showalter}]{Brown2003}
Brown, R.~H., Baines, K.~H., Bellucci, G., {et~al.} 2003, Icarus, 164, 461

\bibitem[{Buratti \& Veverka(1983)}]{Buratti1983}
Buratti, B., \& Veverka, J. 1983, Icarus, 55, 93

\bibitem[{Buratti(1991)}]{Buratti1991}
Buratti, B.~J. 1991, Icarus, 92, 312

\bibitem[{Buratti(1995)}]{Buratti1995}
---. 1995, Journal of Geophysical Research, 100, 19061.
\newblock \url{http://doi.wiley.com/10.1029/95JE00146}

\bibitem[{Cahoy {et~al.}(2010)Cahoy, Marley, \& Fortney}]{Cahoy2010}
Cahoy, K.~L., Marley, M.~S., \& Fortney, J.~J. 2010, The Astrophysical Journal,
  724, 189

\bibitem[{Cowan \& Agol(2008)}]{Cowan2008}
Cowan, N.~B., \& Agol, E. 2008, The Astrophysical Journal, 678, L129.
\newblock \url{http://arxiv.org/abs/0803.3622 http://dx.doi.org/10.1086/588553}

\bibitem[{Cowan \& Agol(2011)}]{Cowan2011}
---. 2011, Astrophysical Journal, 726, doi:10.1088/0004-637X/726/2/82

\bibitem[{Cowan \& Fujii(2017)}]{Cowan2017a}
Cowan, N.~B., \& Fujii, Y. 2017, in Handbook of Exoplanets (Cham: Springer
  International Publishing), 1--16

\bibitem[{Cowan {et~al.}(2009)Cowan, Agol, MeaDows, Robinson, Livengood,
  Deming, Lisse, A'Hearn, Wellnitz, Seager, \& Charbonneau}]{Cowan2009}
Cowan, N.~B., Agol, E., MeaDows, V.~S., {et~al.} 2009, Astrophysical Journal,
  700, 915

\bibitem[{Crow {et~al.}(2011)Crow, McFadden, Robinson, Meadows, Livengood,
  Hewagama, Barry, Deming, Lisse, \& Wellnitz}]{Crow2011}
Crow, C.~A., McFadden, L.~A., Robinson, T., {et~al.} 2011, Astrophysical
  Journal, 729, doi:10.1088/0004-637X/729/2/130

\bibitem[{Dittmann {et~al.}(2017)Dittmann, Irwin, Charbonneau, Bonfils,
  Astudillo-Defru, Haywood, Berta-Thompson, Newton, Rodriguez, Winters, Tan,
  Almenara, Bouchy, Delfosse, Forveille, Lovis, Murgas, Pepe, Santos, Udry,
  W{\"{u}}nsche, Esquerdo, Latham, \& Dressing}]{Dittmann2017}
Dittmann, J.~A., Irwin, J.~M., Charbonneau, D., {et~al.} 2017, Nature, 544, 333

\bibitem[{Domingue \& Verbiscer(1997)}]{Domingue1997}
Domingue, D., \& Verbiscer, A. 1997, Icarus, 128, 49.
\newblock
  \url{http://adsabs.harvard.edu/cgi-bin/nph-data_query?bibcode=1997Icar..128...49D&link_type=EJOURNAL%5Cnpapers3://publication/doi/10.1006/icar.1997.5730}

\bibitem[{Domingue(1998)}]{Domingue1998}
Domingue, D.~L. 1998, Applied Physics, 136, 113

\bibitem[{Dyudina {et~al.}(2016)Dyudina, Zhang, Li, Kopparla, Ingersoll, Dones,
  Verbiscer, \& Yung}]{Dyudina2016}
Dyudina, U., Zhang, X., Li, L., {et~al.} 2016, The Astrophysical Journal, 822,
  76

\bibitem[{Dyudina {et~al.}(2005)Dyudina, Sackett, Bayliss, Seager, Porco,
  Throop, \& Dones}]{Dyudina2005}
Dyudina, U.~A., Sackett, P.~D., Bayliss, D. D.~R., {et~al.} 2005, The
  Astrophysical Journal, 618, 973

\bibitem[{Fan {et~al.}(2019)Fan, Li, Li, Bartlett, Jiang, Natraj, Crisp, \&
  Yung}]{Fan2019}
Fan, S., Li, C., Li, J.-Z., {et~al.} 2019.
\newblock \url{http://arxiv.org/abs/1908.04350}

\bibitem[{Fanale {et~al.}(1974)Fanale, Johnson, \& Matson}]{Fanale1974}
Fanale, F.~P., Johnson, T.~V., \& Matson, D.~L. 1974, Science, 186, 922.
\newblock \url{http://www.sciencemag.org/cgi/doi/10.1126/science.186.4167.922}

\bibitem[{Fujii {et~al.}(2011)Fujii, Kawahara, Suto, Fukuda, Nakajima,
  Livengood, \& Turner}]{Fujii2011}
Fujii, Y., Kawahara, H., Suto, Y., {et~al.} 2011, Astrophysical Journal, 738,
  doi:10.1088/0004-637X/738/2/184

\bibitem[{Fujii {et~al.}(2010)Fujii, Kawahara, Suto, Taruya, Fukuda, Nakajima,
  \& Turner}]{Fujii2010}
---. 2010, Astrophysical Journal, 715, 866

\bibitem[{Fujii {et~al.}(2014)Fujii, Kimura, Dohm, \& Ohtake}]{Fujii2014}
Fujii, Y., Kimura, J., Dohm, J., \& Ohtake, M. 2014, Astrobiology, 14, 753.
\newblock \url{http://online.liebertpub.com/doi/abs/10.1089/ast.2014.1165}

\bibitem[{Fujii {et~al.}(2017)Fujii, Lustig-Yaeger, \& Cowan}]{Fujii2017}
Fujii, Y., Lustig-Yaeger, J., \& Cowan, N.~B. 2017, The Astronomical Journal,
  154, 189.
\newblock
  \url{http://arxiv.org/abs/1708.04886%0Ahttp://dx.doi.org/10.3847/1538-3881/aa89f1}

\bibitem[{Fujii {et~al.}(2018)Fujii, Angerhausen, Deitrick, Domagal-Goldman,
  Grenfell, Hori, Kane, Pall{\'{e}}, Rauer, Siegler, Stapelfeldt, \&
  Stevenson}]{Fujii2018}
Fujii, Y., Angerhausen, D., Deitrick, R., {et~al.} 2018, Astrobiology, 18, 739.
\newblock \url{http://arxiv.org/abs/1705.07098
  http://www.liebertpub.com/doi/10.1089/ast.2017.1733}

\bibitem[{Gaudi {et~al.}(2018)Gaudi, Seager, Mennesson, Kiessling, Warfield,
  Kuan, Cahoy, Clarke, Domagal-Goldman, Feinberg, Guyon, Kasdin, Mawet,
  Robinson, Rogers, Scowen, Somerville, Stapelfeldt, Stark, Stern, Turnbull,
  Martin, Alvarez-Salazar, Amini, Arnold, Balasubramanian, Baysinger, Blais,
  Brooks, Calvet, Cormarkovic, Cox, Danner, Davis, Dorsett, Effinger, Eng,
  Garcia, Gaskin, Harris, Howe, Knight, Krist, Levine, Li, Lisman, Mandic,
  Marchen, Marrese-Reading, McGowen, Miyaguchi, Morgan, Nemati, Nikzad, Nissen,
  Novicki, Perrine, Redding, Richards, Rud, Scharf, Serabyn, Shaklan, Smith,
  Stahl, Stahl, Tang, Van~Buren, Villalvazo, Warwick, Webb, Wofford, Woo, Wood,
  Ziemer, Douglas, Faramaz, Hildebrandt, Meshkat, Plavchan, Ruane, \&
  Turner}]{HabEx}
Gaudi, B.~S., Seager, S., Mennesson, B., {et~al.} 2018, ArXiv.
\newblock \url{http://arxiv.org/abs/1809.09674}

\bibitem[{Geissler {et~al.}(1999)Geissler, McEwen, Keszthelyi, Lopes-Gautier,
  Granahan, \& Simonelli}]{Geissler1999}
Geissler, P., McEwen, A., Keszthelyi, L., {et~al.} 1999, Icarus, 140, 265.
\newblock \url{http://linkinghub.elsevier.com/retrieve/pii/S0019103599961286}

\bibitem[{Gillon {et~al.}(2017)Gillon, Triaud, Demory, Jehin, Agol, Deck,
  Lederer, de~Wit, Burdanov, Ingalls, Bolmont, Leconte, Raymond, Selsis,
  Turbet, Barkaoui, Burgasser, Burleigh, Carey, Chaushev, Copperwheat, Delrez,
  Fernandes, Holdsworth, Kotze, Van~Grootel, Almleaky, Benkhaldoun, Magain, \&
  Queloz}]{Gillon2017}
Gillon, M., Triaud, A. H. M.~J., Demory, B.-O., {et~al.} 2017, Nature, 542, 456

\bibitem[{Grimm {et~al.}(2018)Grimm, Demory, Gillon, Dorn, Agol, Burdanov,
  Delrez, Sestovic, Triaud, Turbet, Bolmont, Caldas, Wit, Jehin, Leconte,
  Raymond, Grootel, Burgasser, Carey, Fabrycky, Heng, Hernandez, Ingalls,
  Lederer, Selsis, \& Queloz}]{Grimm2018}
Grimm, S.~L., Demory, B.-O., Gillon, M., {et~al.} 2018, Astronomy {\&}
  Astrophysics, 613, A68.
\newblock \url{https://www.aanda.org/10.1051/0004-6361/201732233}

\bibitem[{Hegde \& Kaltenegger(2013)}]{Hegde2013}
Hegde, S., \& Kaltenegger, L. 2013, Astrobiology, 13, 47.
\newblock \url{http://arxiv.org/ftp/arxiv/papers/1209/1209.4098.pdf
  http://www.ncbi.nlm.nih.gov/pubmed/23252379
  http://online.liebertpub.com/doi/abs/10.1089/ast.2012.0849}

\bibitem[{Hendrix {et~al.}(2005)Hendrix, Domingue, \& King}]{Hendrix2005}
Hendrix, A.~R., Domingue, D.~L., \& King, K. 2005, Icarus, 173, 29

\bibitem[{Hunter(2007)}]{matplotlib}
Hunter, J.~D. 2007, Computing in Science {\&} Engineering, 9, 90.
\newblock \url{http://ieeexplore.ieee.org/document/4160265/}

\bibitem[{Jiang {et~al.}(2018)Jiang, Zhai, Herman, Zhai, Hu, Su, Natraj, Li,
  Xu, \& Yung}]{Jiang2018}
Jiang, J.~H., Zhai, A.~J., Herman, J., {et~al.} 2018, The Astronomical Journal,
  156, 26.
\newblock
  \url{http://stacks.iop.org/1538-3881/156/i=1/a=26?key=crossref.4c689e82336e83e4773c167ea4efb2c6}

\bibitem[{Jones {et~al.}(2001)Jones, Oliphant, Peterson, \& {others}}]{SciPy}
Jones, E., Oliphant, T., Peterson, P., \& {others}. 2001, {{\{}SciPy{\}}: Open
  source scientific tools for {\{}Python{\}}}, , .
\newblock \url{http://www.scipy.org/}

\bibitem[{Karkoschka(1994)}]{Karkoschka1994}
Karkoschka, E. 1994, Icarus, 111, 174

\bibitem[{Karkoschka(1998)}]{Karkoschka1998}
---. 1998, Icarus, 133, 134

\bibitem[{Knowles(2016)}]{Knowles2016}
Knowles, B. 2016, {Cassini Imaging Science Subsystem (ISS) Data User's Guide},
  PDS.
\newblock \url{https://pds-rings.seti.org/cassini/iss/}

\bibitem[{Knutson {et~al.}(2007)Knutson, Charbonneau, Allen, Fortney, Agol,
  Cowan, Showman, Cooper, \& Megeath}]{Knutson2007}
Knutson, H.~A., Charbonneau, D., Allen, L.~E., {et~al.} 2007, Nature, 447, 183.
\newblock \url{http://www.nature.com/articles/nature05782}

\bibitem[{LEBIGOT(2016)}]{uncertainties}
LEBIGOT, E.~O. 2016, {Uncertainties: a Python package for calculations with
  uncertainties}, , .
\newblock \url{http://pythonhosted.org/uncertainties/}

\bibitem[{Livengood {et~al.}(2011)Livengood, Deming, A'Hearn, Charbonneau,
  Hewagama, Lisse, McFadden, Meadows, Robinson, Seager, \&
  Wellnitz}]{Livengood2011}
Livengood, T.~A., Deming, L.~D., A'Hearn, M.~F., {et~al.} 2011, Astrobiology,
  11, 907.
\newblock \url{http://www.liebertonline.com/doi/abs/10.1089/ast.2011.0614}

\bibitem[{Luger {et~al.}(2018)Luger, Agol, Foreman-Mackey, Fleming,
  Lustig-Yaeger, \& Deitrick}]{STARRY}
Luger, R., Agol, E., Foreman-Mackey, D., {et~al.} 2018

\bibitem[{Luger {et~al.}(2019)Luger, Bedell, Vanderspek, \& Burke}]{Luger2019}
Luger, R., Bedell, M., Vanderspek, R., \& Burke, C.~J. 2019, 1.
\newblock \url{http://arxiv.org/abs/1903.12182}

\bibitem[{Madden \& Kaltenegger(2018)}]{Madden2018}
Madden, J., \& Kaltenegger, L. 2018, Astrobiology, 18, 1559

\bibitem[{Mayorga {et~al.}(2016)Mayorga, Jackiewicz, Rages, West, Knowles,
  Lewis, \& Marley}]{Mayorga2016}
Mayorga, L.~C., Jackiewicz, J., Rages, K., {et~al.} 2016, The Astronomical
  Journal, 152, 209

\bibitem[{McCord {et~al.}(2004)McCord, Coradini, Hibbitts, Capaccioni, Hansen,
  Filacchione, Clark, Cerroni, Brown, Baines, Bellucci, Bibring, Buratti,
  Bussoletti, Combes, Cruikshank, Drossart, Formisano, Jaumann, Langevin,
  Matson, Nelson, Nicholson, Sicardy, \& Sotin}]{McCord2004}
McCord, T.~B., Coradini, A., Hibbitts, C.~A., {et~al.} 2004, Icarus, 172, 104

\bibitem[{McEwen {et~al.}(1985)McEwen, Matson, Johnson, \&
  Soderblom}]{McEwen1985}
McEwen, A.~S., Matson, D.~L., Johnson, T.~V., \& Soderblom, L.~A. 1985, Journal
  of Geophysical Research, 90, 12345

\bibitem[{McKinney(2010)}]{Pandas}
McKinney, W. 2010, in Proceedings of the 9th Python in Science Conference, ed.
  S.~van~der Walt \& J.~Millman, 51--56

\bibitem[{McREYNOLDS \& BLYTHE(2005)}]{McREYNOLDS2005}
McREYNOLDS, T., \& BLYTHE, D. 2005, in Advanced Graphics Programming Using
  OpenGL (Elsevier), 35--56.
\newblock
  \url{https://linkinghub.elsevier.com/retrieve/pii/B9781558606593500056}

\bibitem[{Ment {et~al.}(2019)Ment, Dittmann, Astudillo-Defru, Charbonneau,
  Irwin, Bonfils, Murgas, Almenara, Forveille, Agol, Ballard, Berta-Thompson,
  Bouchy, Cloutier, Delfosse, Doyon, Dressing, Esquerdo, Haywood, Kipping,
  Latham, Lovis, Newton, Pepe, Rodriguez, Santos, Tan, Udry, Winters, \&
  W{\"{u}}nsche}]{Ment2019}
Ment, K., Dittmann, J.~A., Astudillo-Defru, N., {et~al.} 2019, The Astronomical
  Journal, 157, 32

\bibitem[{Millis \& Thompson(1975)}]{Millis1975}
Millis, R., \& Thompson, D.~T. 1975, Icarus, 26, 408.
\newblock \url{http://linkinghub.elsevier.com/retrieve/pii/0019103575901086}

\bibitem[{{National Academies of Sciences, Engineering, and
  Medicine}(2018)}]{Strategy2018}
{National Academies of Sciences, Engineering, and Medicine}. 2018, {Exoplanet
  Science Strategy} (Washington, D.C.: National Academies Press),
  doi:10.17226/25187.
\newblock \url{https://www.nap.edu/catalog/25187}

\bibitem[{Porco {et~al.}(2004)Porco, West, Squyres, Mcewen, Thomas, Murray,
  Delgenio, Ingersoll, Johnson, Neukum, Veverka, Dones, Brahic, Burns,
  Haemmerle, Knowles, Dawson, Roatsch, Beurle, \& Owen}]{Porco2004}
Porco, C.~C., West, R.~a., Squyres, S., {et~al.} 2004, Space Science Reviews,
  115, 363

\bibitem[{Roberge \& {The LUVOIR Team}(2018)}]{LUVOIRInterim}
Roberge, A., \& {The LUVOIR Team}. 2018, ArXiv.
\newblock \url{http://arxiv.org/abs/1809.09668}

\bibitem[{Robinson {et~al.}(2010)Robinson, Meadows, \& Crisp}]{Robinson2010}
Robinson, T.~D., Meadows, V.~S., \& Crisp, D. 2010, The Astrophysical Journal,
  721, L67.
\newblock
  \url{http://stacks.iop.org/2041-8205/721/i=1/a=L67?key=crossref.b66d93f984635c8fc102046d6ae35d31}

\bibitem[{Robinson {et~al.}(2011)Robinson, Meadows, Crisp, Deming, A'Hearn,
  Charbonneau, Livengood, Seager, Barry, Hearty, Hewagama, Lisse, McFadden, \&
  Wellnitz}]{Robinson2011}
Robinson, T.~D., Meadows, V.~S., Crisp, D., {et~al.} 2011, Astrobiology, 11,
  393.
\newblock \url{http://www.liebertonline.com/doi/abs/10.1089/ast.2011.0642}

\bibitem[{Robitaille {et~al.}(2013)Robitaille, Tollerud, Greenfield,
  Droettboom, Bray, Aldcroft, Davis, Ginsburg, Price-Whelan, Kerzendorf,
  Conley, Crighton, Barbary, Muna, Ferguson, Grollier, Parikh, Nair,
  G{\"{u}}nther, Deil, Woillez, Conseil, Kramer, Turner, Singer, Fox, Weaver,
  Zabalza, Edwards, Azalee~Bostroem, Burke, Casey, Crawford, Dencheva, Ely,
  Jenness, Labrie, Lim, Pierfederici, Pontzen, Ptak, Refsdal, Servillat, \&
  Streicher}]{Astropy}
Robitaille, T.~P., Tollerud, E.~J., Greenfield, P., {et~al.} 2013, Astronomy
  {\&} Astrophysics, 558, A33

\bibitem[{{SETI} {et~al.}(2014){SETI}, Showalter, \& French}]{vicar}
{SETI}, Showalter, M., \& French, R. 2014, {vicar},  GitHub.
\newblock \url{https://github.com/SETI/pds-tools/blob/master/vicar.py}

\bibitem[{Simonelli(1994)}]{Simonelli1994}
Simonelli, D. 1994, Icarus, 107, 375.
\newblock
  \url{http://linkinghub.elsevier.com/retrieve/doi/10.1006/icar.1994.1030}

\bibitem[{Simonelli {et~al.}(2001)Simonelli, Dodd, \& Veverka}]{Simonelli2001}
Simonelli, D.~P., Dodd, C., \& Veverka, J. 2001, Journal of Geophysical
  Research: Planets, 106, 33241

\bibitem[{Simonelli \& Veverka(1984)}]{Simonelli1984}
Simonelli, D.~P., \& Veverka, J. 1984, Icarus, 59, 406

\bibitem[{Simonelli \& Veverka(1986{\natexlab{a}})}]{Simonelli1986a}
---. 1986{\natexlab{a}}, Icarus, 66, 403

\bibitem[{Simonelli \& Veverka(1986{\natexlab{b}})}]{Simonelli1986b}
---. 1986{\natexlab{b}}, Icarus, 66, 428

\bibitem[{Simonelli \& Veverka(1986{\natexlab{c}})}]{Simonelli1986c}
---. 1986{\natexlab{c}}, Icarus, 68, 503

\bibitem[{Simonelli \& Veverka(1988)}]{Simonelli1988}
---. 1988, Icarus, 74, 240

\bibitem[{Spencer {et~al.}(1995)Spencer, Calvin, \& Person}]{Spencer1995}
Spencer, J.~R., Calvin, W.~M., \& Person, M.~J. 1995, Journal of Geophysical
  Research, 100, 19049

\bibitem[{{The LUVOIR Team}(2019)}]{TheLUVOIRTeam2019}
{The LUVOIR Team}. 2019.
\newblock \url{http://arxiv.org/abs/1912.06219}

\bibitem[{Thorngren(2019)}]{PlanetSlicer}
Thorngren, D.~P. 2019, {PlanetSlicer},  GitHub.
\newblock \url{https://github.com/dpthorngren/PlanetSlicer}

\bibitem[{Travis~E(2006)}]{NumPy}
Travis~E, O. 2006, {A guide to NumPy},  USA: Trelgol Publishing

\bibitem[{{US Geological Survey}(2001)}]{CallistoMap}
{US Geological Survey}. 2001, U.S. Geological Survey Geologic Investigations
  Series, I-2770.
\newblock \url{https://pubs.usgs.gov/imap/2770/}

\bibitem[{{US Geological Survey}(2002)}]{EuropaMap}
---. 2002, U.S. Geological Survey Geologic Investigations Series, I–2757.
\newblock \url{https://pubs.usgs.gov/imap/i2757/}

\bibitem[{{US Geological Survey}(2003)}]{GanymedeMap}
---. 2003, U.S. Geological Survey Geologic Investigations Series, I-2762.
\newblock \url{https://pubs.usgs.gov/imap/i2762/}

\bibitem[{van~der Walt {et~al.}(2014)van~der Walt, Sch{\"{o}}nberger,
  Nunez-Iglesias, Boulogne, Warner, Yager, Gouillart, \& Yu}]{skimage}
van~der Walt, S., Sch{\"{o}}nberger, J.~L., Nunez-Iglesias, J., {et~al.} 2014,
  PeerJ, 2, e453.
\newblock \url{https://peerj.com/articles/453}

\bibitem[{West {et~al.}(2010)West, Knowles, Birath, Charnoz, Di~Nino, Hedman,
  Helfenstein, McEwen, Perry, Porco, Salmon, Throop, \& Wilson}]{West2010}
West, R., Knowles, B., Birath, E., {et~al.} 2010, Planetary and Space Science,
  58, 1475

\bibitem[{Williams {et~al.}(2011)Williams, Keszthelyi, Crown, Yff, Jaeger,
  Schenk, Geissler, \& Becker}]{Williams2011}
Williams, D.~A., Keszthelyi, L.~P., Crown, D.~A., {et~al.} 2011, Icarus, 214,
  91

\end{thebibliography}
\end{document}